\documentclass{aa}  
\usepackage{graphicx}
\usepackage[varg]{txfonts}
\usepackage[colorlinks=true,citecolor=blue]{hyperref}
\usepackage{natbib}
\usepackage{lscape}
\usepackage{multirow}
\usepackage{multicol}
\usepackage{supertabular}
\usepackage{etoolbox}

\newcommand{\micron}{\,\mu\mathrm{m}}
\newcommand{\microJybm}{\,\mathrm{\mu Jy\,beam^{-1}}}
\newcommand{\microJy}{\,\mathrm{\mu Jy}}
\newcommand{\ergs}{\,\mathrm{erg\,s^{-1}}}
\newcommand{\vega}{_{V}}
\newcommand{\ab}{_\mathrm{AB}}

\begin{document}

	\title{Nowhere to Hide: Radio-faint AGN in the GOODS-N field}
	
	\subtitle{II. Multi-wavelength AGN selection techniques and host galaxy properties}
	
	\author{J. F.~Radcliffe\inst{1,2,3}
		\and P. D.~Barthel\inst{1}
		\and A. P.~Thomson\inst{3}
		\and M. A.~Garrett\inst{3,4} 
		\and R. J.~Beswick\inst{3}
		\and T. W. B.~Muxlow\inst{3}
	}
	
	\institute{Kapteyn Astronomical Institute, University of Groningen, 9747 AD Groningen, The Netherlands \\
		\email{jack.radcliffe@up.ac.za}
		\and Department of Physics, University of Pretoria, Lynnwood Road, Hatfield, Pretoria, 0083, South Africa
		\and Jodrell Bank Centre for Astrophysics, School of Physics \& Astronomy, The University of Manchester, Alan Turing Building, Oxford Road, Manchester M13 9PL, UK
		\and Leiden Observatory, Leiden University, PO Box 9513, 2300 RA Leiden, The Netherlands}
	
	\date{Received <date> / Accepted <date>}
	
	\titlerunning{Nowhere to Hide: Radio-faint AGN in the GOODS-N field. II.}
	\authorrunning{J. F. Radcliffe et al.}

\abstract
{Obtaining a census of active galactic nuclei (AGN) activity across cosmic time is critical to our understanding of galaxy evolution and formation. Many AGN classification techniques are compromised by dust obscuration. However, very long baseline interferometry (VLBI) can be used to identify high brightness temperature compact radio emission ($>10^{5}\,\mathrm{K}$) in distant galaxies that can only be reliably attributed to AGN activity.}
{We present the second in a series of papers dealing with the compact radio population in the GOODS-N field. This paper reviews the various multi-wavelength data and AGN classification techniques in the context of a VLBI-detected sample and use these to investigate the nature of the AGN as well as their host galaxies.} 
{Multi-wavelength data from radio to X-ray were compiled for the GOODS-N AGN sample, and fourteen widely used multi-wavelength AGN classification schemes were tested. We discuss and compare the various biases that affect multi-wavelength and VLBI selection. We use the physical interpretation to imply the nature of VLBI-selected AGN and their hosts.} {Firstly, we find that no single identification technique can identify all VLBI objects as AGN. Infrared colour-colour selection is most notably incomplete. However, the usage of multiple classification schemes can identify all VLBI-selected AGN, independently verifying similar approaches used in other deep field surveys. In the era of large area surveys with instruments such as the SKA and ngVLA, multi-wavelength coverage, which relies heavily upon observations from space, is often unavailable. Therefore, VLBI remains an integral component in detecting AGN of the jetted efficient and inefficient accretion types. Secondly, a substantial fraction (46\%) of the VLBI AGN have no X-ray counterpart, which is most likely due to lack of sensitivity in the X-ray band. Thirdly, a high fraction of the VLBI AGN reside in low or intermediate redshift dust-poor early-type galaxies. These most likely exhibit inefficient accretion. Fourthly, A significant fraction of the VLBI AGN reside in symbiotic dusty starburst - AGN systems. Finally, in the Appendix, we present an extensive compilation of the multi-wavelength properties of all the VLBI AGN in GOODS-N.}
{}
        
\keywords{radio continuum: galaxies - galaxies: active - techniques: high angular resolution, interferometric}

\maketitle

\defcitealias{LacyIRAC2007}{L07}
\defcitealias{donley2012identifying}{D12}
\defcitealias{Kirkpatrick2012IR}{K12}
\defcitealias{muxlow2005high}{M05}
\defcitealias{stern2005mid}{S05}
\defcitealias{Jarrett2011:WISE}{J11}
\defcitealias{Stern2012:ir}{S12}
\defcitealias{Mateos2012:irx}{M12}
\defcitealias{Radcliffe2018:p1}{Paper I}

\section{Introduction}\label{Sec: P2-Introduction}

Deep, wide-field surveys of the sky have yielded a profound understanding as to the evolution of galaxies. Surveys at longer wavelengths, especially radio and far-infrared (FIR), play a crucial role here as dust and its attendant obscuration is a ubiquitous partner to the merger activity associated with galaxy growth \citep[e.g.][]{Zinn2011}. It is believed that minor mergers and/or cold gas accretion, rather than major mergers, are responsible for the growth of these systems \citep[e.g.][]{Elbaz2011}. Given the well-established scaling relations, super-massive black holes (SMBH) must have been in place early and their episodic growth must manifest through accretion-related radiation.

A widespread symbiotic occurrence of star formation and super-massive black hole (SMBH) growth at high redshifts is expected. Indeed, this is seen in radio and FIR observations of faint X-ray selected active galactic nuclei (AGN) \citep[e.g.][]{Padovani2009,Mullaney2012,Rodighiero2015} and in radio-loud AGN \citep[e.g.][]{Podigachoski2015}. Recent literature has reached a consensus that star-formation (SF) and SMBH accretion were more common in the past and peak at redshifts of around 2 \citep[see][and references therein]{Madau:2014gt}.

There are at least two important issues that still need to be addressed: first, the nature of galaxies with dust-obscured AGN and second, the interplay between nuclear activity and star formation. To achieve this we require a complete census of AGN activity. X-ray surveys have proved to be a particularly powerful method of selecting both obscured and un-obscured AGN to faint flux densities and high redshifts. However, Compton-thick AGN, where the X-ray emission below $10\,\mathrm{keV}$ are attenuated by obscuration with column densities larger than $5\times10^{24}\,\mathrm{cm^{-2}}$ have been routinely missed by these surveys \citep[e.g.][]{hasinger2008absorption}. A recent study by \citet{Mateos:2017fm}, using \textit{XMM-Newton}, predicted that Compton-thick AGN may account for as much as $37\substack{+9 \\ -10}\%$ of the total AGN population, and so the majority of luminous accreting black holes at $z<1$ are so embedded that they remain undetected by current wide-area X-ray surveys. Synthesis modelling of the X-ray background (XRB) seems to confirm this, revealing the need for a large population of heavily-obscured AGN in order to replicate the high energy peak ($\sim 30$\,keV) seen in the unresolved XRB \citep{GilliX_ray2007,Ballantyne:2011em}.

With the operational capabilities provided by the \textit{Spitzer} and \textit{Herschel} telescopes, considerable efforts have been made to identify obscured AGN activity by using the infrared (IR) bands. In the mid-IR (MIR), a typical star-forming galaxy has a dip in the IR spectral energy distribution (SED) between the long wavelength emission from star formation heated dust (typically $\lambda\sim100\micron$ with dust temperatures of $25\mbox{-}50\,\mathrm{K}$) and the $1.6\micron$ stellar bump. If an AGN is present, dust in the torus surrounding the central black hole will be heated to $200\mbox{-}1500\,\mathrm{K}$ due to the absorption of ultra-violet (UV) photons. These are then emitted into the MIR bands \citep[$5\mbox{-}40\micron$; e.g.][]{feltre2013}. This can result in a flattening of IR SED between SF heated dust and stellar emission \citep[e.g.][]{mullaney2011defining,donley2012identifying}. IR surveys have tried to identify this hot AGN-related dust component from stellar emission using various methods, such as \textit{Spitzer} IRAC power law emission fitting \citep{alonso2006infrared}, \textit{Spitzer} IRAC/MIPS colour-colour diagnostics \citep[e.g.][]{Lacy2004:ir,LacyIRAC2007,donley2012identifying,Kirkpatrick2012IR}, and WISE colour-colour diagnostics \citep[e.g.][]{Jarrett2011:WISE,Stern2012:ir,Mateos2012:irx}. However, it has been noted that SF related emission can dominate across the entire IR band that can mask the presence of AGN-induced hot dust \citep[e.g.][]{2003MNRAS.343..585F}.

A potentially powerful approach comes in the form of radio observations, that provide a dust-independent window into the obscured and un-obscured AGN populations. At high flux densities ($>\,\mathrm{mJy}$), the radio population is dominated by the powerful radio-loud population of large ($\gg\,\mathrm{kpc}$) extended radio galaxies and quasars powered by AGN. Towards fainter flux densities ($<\,\mathrm{mJy}$) the radio population transitions into a dominant population of star-forming galaxies and `non-jetted' AGN, whose radio emission is often confined to a compact core \citep[see][and references therein]{Padovani2016}. These pose a problem because the majority of deep radio surveys are conducted at low ($\sim$ arcsecond) resolutions, corresponding to $>\,\mathrm{kpc}$ scales in distant galaxies. As a result, synchrotron emission from AGN activity is often merged with SF-related emission. To disentangle these contributions and to isolate the AGN-related radio emission, various approaches can be used.

The first is to use the well known FIR radio correlation (FIRRC). This exists in star-forming galaxies because radio emission is intimately correlated with the FIR due to their mutual origin in active high-mass star-forming regions. This correlation is found to hold at high redshifts  \citep{yun2001radio,Garrett:2002ft,Sargent2010:co,thomson14,Pannella:2015gk,Magnelli2015:firrc,Delhaize2017:firrc,algera20firrc}, thus any AGN activity can be identified by deviations from this correlation. Here the AGN produces the excess radio emission \citep[e.g.][]{Donley:2005da,Delmoro2013}. In the absence of reliable FIR observations, MIR bands such as the {\it Spitzer} MIPS $24\,\mu\mathrm{m}$ can be used as a proxy to great effect \citep[e.g.][]{Appleton24um2004, chi2013deep}. Again though this method {\it only} works when the AGN is the dominant source of the total radio emission, so any weak embedded AGN can be hidden behind radio emission originating from stellar processes.

A possible solution is to use Very Long Baseline Interferometry (VLBI). The sparsity of a VLBI array means that it is only sensitive to compact,`point-like', radio sources that have brightness temperatures in excess of $10^5\,\mathrm{K}$. Even the most luminous starburst galaxies have brightness temperatures less than $10^5\,\mathrm{K}$ meaning that a VLBI detection is a reliable indicator of AGN activity at high redshift \citep[e.g.][]{condon1982temperatures,kewley2000compact,2011A&A...526A..74M}. In recent years, technological developments have permitted degrees of the sky to be surveyed at high resolution, thus the use VLBI as a dust-independent tracer of AGN activity is now finally possible.

This study, the second in a series dealing with the ultra-faint radio population in the Great Observatories Origins Deep Survey-North field \citep[GOODS-N;][]{2004ApJ...600L..93G}, aims to use a VLBI-selected sample of AGN to compare, contrast and test multi-wavelength AGN classification techniques. These shall be used to infer the nature of VLBI-selected AGN.

The paper is organised as follows. In Section~\ref{Sec:P2-Data}, we introduce the various multi-wavelength data and catalogues. We investigate the performance of multiple AGN selection techniques in the context of our VLBI sample ranging from the radio to X-ray in Section~\ref{Sec:AGN_class_techniques}. In Section~\ref{Sec:AGN_overlap}, we compare the various classification methods, including their biases and limitations, and we infer what these imply about the nature of VLBI-AGN. We summarise our findings in Section~\ref{Sec:P2-conclusions}, and provide detailed descriptions of the individual VLBI-selected AGN and their hosts in Appendix\,\ref{Appendix}.

Throughout the paper we use the following standards. 1. A spatially-flat 6-parameter $\mathrm{\Lambda CDM}$ cosmology with $H_0 = 67.8\pm0.9\,\mathrm{km\,s^{-1}}$ $\mathrm{Mpc^{-1}}$, $\Omega_\mathrm{m} = 0.308\pm0.012$ and $\Omega_{\Lambda} = 0.692 \pm 0.012$ \citep{Planck2016}. 2. The convention $S_\nu \propto \nu^{\alpha}$, where $S_\nu$ is the integrated flux density and $\alpha$ is the intrinsic source spectral index. 3. The subscripts $\vega$ and $_\mathrm{AB}$ correspond to the $Vega$ and AB magnitude systems, respectively.

\section{Observations, data, and catalogues}\label{Sec:P2-Data}

\begin{figure*}
	\centering
	\includegraphics[width=0.85\hsize]{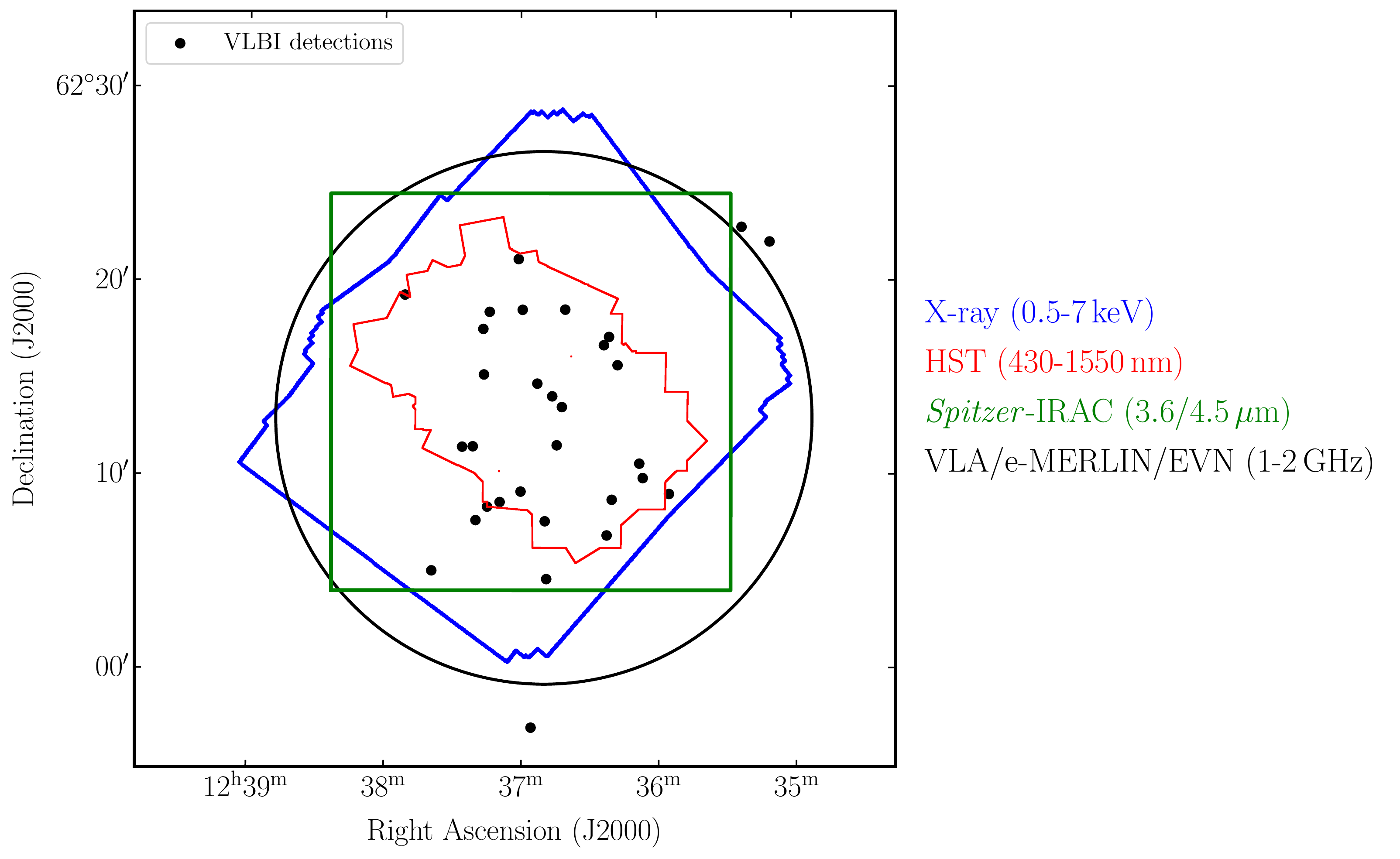}
	\caption{Sub-set of observations in the GOODS-N field illustrating the multi-wavelength coverage. VLBI detections from \citetalias{Radcliffe2018:p1} are plotted as black circles, and the field-of-view (FoV) of the multiple surveys are colour-coded. The non-uniform coverage across multiple instruments is clearly shown. The FoVs were calculated using an edge detection algorithm on the exposure maps apart from radio observations, which corresponds to the HPBW of a 25m telescope at $1.5\,{\rm GHz}$ ($\sim27\farcm5$). }
	\label{Fig:multi-wave_coverage}
\end{figure*}
\begin{table*}
	\caption{Multi-wavelength data available on the GOODS-N field used in these analyses.}             
	\label{table:data_used}      
	\centering  
	{\footnotesize
	\begin{tabular}{llll}
		\hline\hline 
		Band / Filter & Telescope & Survey & Reference(s) \\
		\hline
		$G, R_s$ & Keck/LRIS & & \citet{steidel20013HDFN} \\
		$U, B, V, R, I, z', HK'$& Subaru/Suprime-Cam& Hawaii HDFN & \citet{capak2004HDFN} \\
		$J, H$ & CFHT, UH 2.2m & & \citet{Keenan2010} \\
		$K_{s}$ & CFHT & & \citet{Wang2010} \\
		F140W & HST/WFC3 & 3D-HST & \citet{Skelton_HST3D_2014} \\
		F125W, F160W & HST/WFC3 & CANDELS & \citet{grogin2011CANDELS,koekemoer2011CANDELS} \\
		F435W, F606W, F775W, F850LP & HST/ACS &GOODS & \citet{2004ApJ...600L..93G} \\
		3.6, 4.5 $\mathrm{\mu}$m & \emph{Spitzer}/IRAC & SEDS & \citet{ashby2013IRAC,Wang2010}; \\
        &  & & \citet{Yang_photz_2014} \\
		5.8, 8.0 $\mathrm{\mu}$m & \emph{Spitzer}/IRAC & GOODS & \citet{dickinson2003spitzer,Wang2010} \\
        &  & & \citet{Yang_photz_2014} \\
         3.4, 4.6, 12, 21 $\mathrm{\mu}$m & WISE & AllWISE & \citet{Cutri2012:WISE} \\
		24, 70 $\mathrm{\mu}$m & \emph{Spitzer}/MIPS & GOODS Legacy & \citet{dickinson2003spitzer,Magnelli:2011eb} \\
		100, 160\,$\mathrm{\mu}$m & \emph{Herschel}/PACS & GOODS-\emph{Herschel} & \citet{Elbaz2011}; Thomson et al. (in prep.)\\
		250, 350, 500\,$\mathrm{\mu}$m & \emph{Herschel}/PACS & GOODS-\emph{Herschel} & \citet[][]{Elbaz2011}; Thomson et al. (in prep.)\\
		450, 850 $\mathrm{\mu}$m & SCUBA-2 \& SMA & SUPER GOODS & \citet{Cowie2017:sc} \\
		0.5-7.0 keV & \emph{Chandra} & CDF-N & \citet{xue_xray_2016} \\
		1-2 GHz & VLA & & \citet{morrison2010very}; \citet{Owen2018;gn} \\
		&  & & \citet{2020MNRAS.tmp.1396M}; \citet{Radcliffe2018:var} \\
		& MERLIN-VLA & & \citet{muxlow2005high,richards2000vla} \\
		& $e$-MERLIN-VLA & $e$-MERGE & \citet{2020MNRAS.tmp.1396M} \\
		& EVN & & \citet{garrett2001agn}; \citetalias{Radcliffe2018:p1} \\
		& EVN+VLBA+GBT & & \citet{chi2013deep} \\
		5.5 GHz & VLA & & \citet{Guidetti:2017wt}\\
		8.4 GHz & VLA &  & \citet{richards1998vla} \\
		10 GHz & VLA & & \citet{Murphy:2017ja} \\
		\hline
	\end{tabular}}
\end{table*}

The GOODS-N field covers approximately $160\,\mathrm{arcmin}^2$ and is centred upon the {\it Hubble} Deep Field-North (HDF-N; $12^{h}36^{m}, 62^{\circ}14\arcmin$). The field constitutes some of the deepest multi-wavelength data including \textit{Hubble} Space Telescope (HST), \textit{Chandra}, \textit{Spitzer}, \textit{Herschel}, UBVRIJHK photometry and spectroscopy along with deep radio data from 1-10\,GHz. As a useful guide, the data used in the subsequent analyses is presented in Table~\ref{table:data_used} and the field-of-view (FoV) of a sub-set of these data is presented in Figure~\ref{Fig:multi-wave_coverage}.

\subsection{The VLBI-selected AGN sample}

For completeness, we summarise the VLBI data used in these analyses here but we refer the reader to \citet[][hereafter \citetalias{Radcliffe2018:p1}]{Radcliffe2018:p1} for further details. The GOODS-N field was observed at 1.6\,GHz for 24 hours, using 10 telescopes of the European VLBI Network (EVN). In total, 31 VLBI sources above a $7\sigma$ local r.m.s. were detected within the $0.5^\circ$ FoV. This almost triples the number of VLBI detected sources in this field. The central rms of these observations is approximately $9\microJybm$. The redshifts used are presented in \citetalias{Radcliffe2018:p1}.

For these 31 VLBI detected sources, multi-wavelength data were compiled using a nominal 0\farcs5 search radius for the majority of catalogues. In order to prevent mis-identifications, the false identification rate was calculated using a Monte-Carlo approach. For each multi-wavelength catalogue the following was conducted. Firstly, the coordinates of VLBI sources within the catalogue FoV were randomised within the same FoV. These were then cross-matched with the catalogue using the designated search radius. The total number of false matches is then divided by the total number of coordinates to give a false identification rate for one realisation of randomised coordinates. Finally, this is repeated and the average of the false identification rate was calculated. This was repeated 500 times and the average number of false detections was then divided by the total number of VLBI sources, thus providing the false detection rate. For the majority of catalogues, the $0\farcs5$ search radius gave false detection rates $<1.5\%$. It is worth noting that the GOODS-N field does not have uniform multi-wavelength coverage across the FoV. This means that only sub-sets of VLBI sources can be investigated for each AGN classification technique (e.g. see Figure~\ref{Fig:multi-wave_coverage}). The following subsections outline the various multi-wavelength catalogues derived for our VLBI-selected sources, the results of which are summarised in Table~\ref{table:VLBI_counterparts}.

\subsection{Infrared}\label{sect:deblending}

Near-IR (NIR) and MIR counterparts for the VLBI sources were derived by cross-matching the $K_s$ selected catalogue of \citet{Wang2010}, that includes re-reduced \textit{Spitzer} IRAC photometry (3.6, 4.5, 5.6 and 8$\micron$) to within a $1\arcsec$ radius. The $K_s$ imaging was performed by the WIRCam instrument on the 3.6\,m Canada-France-Hawaii Telescope (CFHT) and covers $0.25\,\mathrm{deg}^2$ down to limiting AB magnitudes of $24.45\,\mathrm{mag}$ ($\sim0.6\microJy$). A total of $30/31$ $K_s$ and $23/31$ \textit{Spitzer} IRAC counterparts were found to within the $1\arcsec$ search radius. To ensure that there is no systematic shift between the two catalogues that could result in false or missing associations, we calculated the median Right Ascension (RA) and Declination (Dec) shift between the VLBI and \citet{Wang2010} catalogue. This was found to be $\Delta\mathrm{RA} = 18.9\pm 69.7\,\mathrm{mas}$ and $\Delta\mathrm{Dec} = -45.2\pm 57.7\,\mathrm{mas}$. While these offsets are much larger than the VLBI beam ($\sim 4\mbox{-}15\,\mathrm{\rm mas}$), they are much smaller than the errors (calculated using the median absolute deviation), and are $<10\%$ of the average seeing ($\sim 0\farcs7\mbox{-}0\farcs8$) of the $K_s$ band observations. We conclude that there is no significant systematic offset between the NIR and VLBI astrometric frames.

We searched for additional counterparts using the \citet{Yang_photz_2014} photometric catalogue, which includes data from the \emph{Spitzer} Extended Deep Survey \citep[SEDS; ][]{ashby2013IRAC}. The \citet{Yang_photz_2014} catalogue is astrometrically aligned to the VLA 1.4\,GHz positions of \citet{morrison2010very} so no systematic adjustments were required. The \citet{Yang_photz_2014} catalogue yielded an additional four IRAC $3.6\micron$ and $4.5\micron$ counterparts and five additional IRAC $5.8\micron$ and $8.0\micron$ counterparts, while the \citet{ashby2013IRAC} catalogue yielded an additional two IRAC $3.6\micron$ and $4.5\micron$ counterparts. In total, there are 30 $K_s$, 27 IRAC $3.6\micron$ and $4.5\micron$, and 28 IRAC $5.8\micron$ and $8.0\micron$ counterparts.

For the \emph{Spitzer} MIPS $24\micron$ fluxes, we cross-matched the VLBI positions with catalogue released with the GOODS survey \citep{dickinson2003spitzer} to within $1\arcsec$, finding 20 counterparts and 4 upper limits. Again, we checked that there was no astrometric offset between the two catalogues ($\Delta\mathrm{RA} = 26.58\pm 95.80\,\mathrm{mas}$, $\Delta\mathrm{Dec} = -17.39\pm 63.30\,\mathrm{mas}$). Additional counterparts were searched for in the \citet{Magnelli:2011eb} catalogue with no additional $24\micron$ detections found. For each IR AGN classification scheme, we impose the condition that they must be clear detections (i.e. $S_\nu>3\sigma$) for all bands used in the scheme.

Due to the inhomogeneous coverage provided by \textit{Spitzer}, further IR counterparts were compiled using the data from the all-sky survey performed by the \textit{Wide-field Infrared Survey Explorer} \citep[WISE;][]{Wright2010:wise,Cutri2012:WISE}. The entire sky at $3.4$, $4.6$, $12$ and $21\, \micron$ was surveyed to 5$\sigma$ point source sensitivities of at least $0.08$, $0.11$, $1$ and $6\,\mathrm{mJy}$, respectively. In order to compare to WISE colour-colour selection schemes, we cross-matched the WISE all-sky catalogue \citep{Cutri2012:WISE} to the VLBI catalogue. We found that there was no significant offset between the VLBI astrometry and the WISE positions ($\Delta {\rm RA} = -0.5\pm 132.7\,{\rm mas}$ and $\Delta {\rm Dec} = 45.7\pm 105.31\,{\rm mas}$. Despite the large beam size of WISE ($6\farcs1$ at $3.4\micron$), we used a small $0\farcs5$ cross-matching radius to ensure that reliable counterparts were found. Using this matching radius, a total of 13 counterparts were found in the WISE $3.6\micron$ and $4.5\micron$ bands, two in the WISE $12\micron$ band and just one in the WISE $22\micron$ band. The difference between the number of detections is because the point source sensitivity is significantly worse in the $12$ and $21\micron$ bands. Due to the small number of VLBI sources, for all the aforementioned IR cross-matching, we visually checked the quoted coordinates to the HST NIR (F160W) \citep{Skelton_HST3D_2014} or the $K_s$ 2$\micron$ images \citep{Wang2010} to ensure that the associations were reliable and no counterparts were missing.

We measured FIR fluxes from \textit{Herschel} PACS and SPIRE imaging of the field, which were undertaken as part of the GOODS-\textit{Herschel} survey \citep{Elbaz2011}. Due to the large \textit{Herschel} beam size ($\sim5\arcsec$ at $100\,\mu$m, rising to $\sim35\arcsec$ at $500\,\mu$m), it is important to account for the effects of source blending before measuring the flux densities of our targets. This well-developed deblending technique has already been used to measure flux densities in confusion-limited \textit{Herschel} SPIRE maps in the Extended \textit{Chandra} Deep Field South \citep[ECDFS;][]{swinbank2014} and Cosmic Evolution Survey \citep[COSMOS;][]{Thomson2017:de} fields. Its extension to the GOODS-N field will be presented in detail in a forthcoming paper (Thomson et al., in prep). For completeness, we briefly summarise the deblending process here. 

Before de-blending, a prior catalogue was compiled containing 3848 source positions provided by $>5\sigma$ \textit{Spitzer} MIPS $24\micron$ detections and $>5\sigma$ detections from the new $1.5\,\mathrm{GHz}$ VLA imaging of GOODS-N \citep{Owen2018;gn,2020MNRAS.tmp.1396M}. We then deblended the \textit{Herschel} PACS and SPIRE maps according to the following process. Firstly, the \textit{Herschel} PACS/SPIRE maps were regridded and resampled to match the astrometry of the VLA 1.5\,GHz continuum image of the field\footnote{The \textit{Herschel} SPIRE maps cover the majority of the VLA and VLBI fields of view, however the PACS images cover only $\sim 25\%$ of the region mapped at longer wavelengths.} \citep{Owen2018;gn,2020MNRAS.tmp.1396M}, and the PACS/SPIRE point spread functions \citep[PSFs;][]{Elbaz2011} were re-sampled to the same pixel scale. Secondly, to check the astrometric alignment of our images, we performed a stacking analysis on the brightest 50 VLA 1.5\,GHz sources in each of the 100, 160, 250, 350 and 500\,$\mu$m images and applied small linear shifts in RA and Dec (in every case $\Delta\leq 1\farcs5$) to bring the stacked \textit{Herschel} peaks in to alignment with the centroid of the stacked radio image. Thirdly, each \textit{Herschel} image was then split in to multiple tiles $1\arcmin\times 1\arcmin$ in size. Fourthly, in each tile in each waveband we created a model image comprised of delta functions located at the positions of galaxies in the prior catalogue with randomly assigned flux densities, and then convolved the model with the appropriate PACS or SPIRE PSF. Fifthly, we randomly perturbed the flux densities in each model image 1000 times before identifying the model which best matches the original image (i.e. has the lowest $\chi^2$). Finally, we use this model as the starting point for the next generation of 1000 models (perturbing the flux densities using the flux density distribution from the previous generation of models) and repeat the process until convergence, that is the generation at which all model images lie within $\Delta\chi^2 = 1\sigma$ of the best-fit.

We define a source as `detected' if its deblended flux density is $\geq3\times$ the un-deblended flux remaining in the residual image (i.e. ${\rm data}-{\rm model}$) at its position. We assign upper-limits to the flux densities of sources that do not meet this criterion by measuring the flux density in the residual image, after the deblending process has subtracted any contribution from neighbouring bright sources. 

In total, 16/31 VLBI-selected sources have measured deblended flux densities at $100\,\mu$m and $160\,\mu$m. 11/31 have upper-limits in these bands, and 4/31 lie outside the \textit{Herschel} PACS survey area. A total of 4/31 sources were found in the SPIRE $250\,\mu$m image (of which 3 and 2 were detected at $350\,\mu$m and $500\,\mu$m, respectively), with 25/31 having upper-limits and 2/31 lying outside the SPIRE survey area.

\subsection{X-rays}

The GOODS-N field has some of the deepest \textit{Chandra} X-ray coverage with full band ($0.5\mbox{--}7\,{\rm keV})$ flux limits of $\sim3.5\times10^{-17}{\rm \,erg\,cm^{-2}\,s^{-1}}$ corresponding to a total exposure time of 2\,Ms. \citet{xue_xray_2016} present a catalogue of 683 X-ray sources detected. These were detected using \texttt{WAVDETECT} with a false positive threshold of less than $10^{-5}$, and a binomial probability source-selection criterion of less than 0.004. This catalogue represents a significant improvement over the previous X-ray catalogue in GOODS-N \citep{Alexander2003}, with an extra 186 sources detected.

We cross matched our VLBI sources to \citet{xue_xray_2016} using a 1\arcsec~radius. In the sample of 31 VLBI-detected sources, 28 sources are located within the \textit{Chandra} exposure (see Figure~\ref{Fig:multi-wave_coverage}). Of the 28 detectable sources, 64\% (18/28) have X-ray counterparts. This fraction is in line with the number mentioned in Section~\ref{Sec: P2-Introduction}.

In order to ensure that the X-ray luminosities are correct, we compared the redshifts used in calculating the luminosities with the VLBI-ascribed redshifts. In total, 16/18 redshifts were within $<1\%$ of each other. For the remaining 2 sources with incorrect redshifts (J123642+621331 and J123714+621826; see Appendix~\ref{Appendix:Previous_detections}), the intrinsic (absorption-corrected) full-band flux, $f_{\mathrm{0.5\mbox{-}7\,keV,int}}$, was recalculated. We followed the steps outlined in Section~3.4 of \citet{XueXray2010} to obtain the full-band X-ray flux, $f_{\mathrm{0.5\mbox{-}7\,keV,int}}$. The absorption-corrected $0.5\mbox{-}7\,\mathrm{keV}$ X-ray luminosity ($L_{\mathrm{0.5\mbox{-}7\,keV}}$) was calculated using the following equation,

\begin{equation}\label{Eqn:X-ray_luminosities}
L_{\mathrm{0.5\mbox{-}7\,keV}} = 4\pi d_{L}^{2}f_{\mathrm{0.5\mbox{-}7\,keV,int}}(1+z)^{\Gamma-2},
\end{equation}

\noindent where $z$ is the source redshift, $d_L$ is its corresponding luminosity distance, and $\Gamma$ is the photon index. For the remaining 10 X-ray undetected VLBI sources, we X-ray luminosity upper limits were derived by replacing $f_{\mathrm{0.5\mbox{-}7\,keV,int}}$ with the full band flux limit at the VLBI position. We assumed an intrinsic photon index of $1.8$, which is typical of an X-ray AGN spectrum \citep[e.g.][]{Tozzi:2006xr}.

\subsection{Radio}

In this study, we use the $1\mbox{-}2\,\mathrm{GHz}$ VLA data reduced as part of the $e$-MERGE survey \citep{2020MNRAS.tmp.1396M}. We refer the reader to this paper for further details. The central rms of the observations was approximately 1.8$\microJybm$ with a median of $3\microJybm$. All VLBI sources had counterparts (within a $0\farcs5$ search radius). The radio luminosities of the VLBI-detected objects range from $\sim 10^{22}\mbox{-}10^{26}\,\mathrm{W\,Hz^{-1}}$ with a median luminosity of $3.5 \times 10^{24}\,\mathrm{W\,Hz^{-1}}$ \citep[see Figure 6 from][]{Radcliffe2018err}.

In order to obtain spectral index measurements, we use the 5.5\,GHz VLA observations of \citet{Guidetti:2017wt}. The VLA observed at 5.5\,GHz in the A- and B-configuration for 14 and 2.5 hours respectively. This produced a 13\farcm5 diameter area with a central rms of $\sim 2\microJybm$. A total of 94 sources were extracted, above a 5$\sigma$ threshold. We refer the reader to \citet{Guidetti:2017wt} for further details.

\subsection{Host morphologies}\label{ssec:host_morph}

The host galaxy morphologies of the VLBI-detected objects were found using the \textit{Hubble} Space Telescope (HST) F125W/F160W images from the 3D-HST survey \citep{Skelton_HST3D_2014}. For those sources outside this area, the ultra-deep Hawaii HDF-N \citep{capak2004HDFN}, or the CFHT $K_s$-band images presented in \citet{Wang2010} are used. Due to the positional accuracy of the VLBI observations, a visual comparison between the optical images and the VLBI positions is usually sufficient for an acquisition of a host galaxy association and its morphological type. Following a similar technique to \citet{2013A&A...551A..97M}, we visually group the hosts into the following morphological types: early-type and bulge dominated, late-type and spiral, irregular morphology or unclassified (i.e. have a low surface brightness or are unresolved).

We define the early type and bulge dominated group as those circular or elliptical extended objects whose surface brightness distribution drops towards the edge. The irregular category encompasses those with clumpy surface brightness distributions. The sources that are unclassified are those where a morphology cannot be attained. This is most likely due to a low signal-to-noise resulting in low surface brightness areas not detected (hence they often appear `point-like').

In addition to these categories, the high-quality and high-resolution afforded from the HST data allow us to check for possible interactions and/or active mergers between the VLBI hosts and the surrounding galaxies which could be influencing AGN and star-formation activity. We define a merger here as having a tidal tail or disturbances to the host or surrounding galaxies, or a secure redshift of any nearby galaxies such that the physical distance between optical nuclei is $< 75\,{\rm kpc}$ \citep[as defined by][]{Larson2016:mr}.\footnote{We note that in \citet{Larson2016:mr} this is a projected separation, whereas we calculated the 3D distance using the available redshift information.}

Optical/NIR counterparts were found to all 31 sources, but six were unresolved or unclassified. Concerning the remaining 25 sources, we find that 72\% (18/25) are hosted in early type systems with 8\% (2/25) hosted in late-type systems. The remaining 16\% (4/25) are hosted in irregular systems. All together 28\% (7/25) show evidence of interactions that are distributed as 4 irregular and 3 early type systems. It is not surprising that these VLBI sources are primarily hosted by elliptical types. The median radio power of the VLBI sample, $3.5\times10^{24}\,\mathrm{W\,Hz^{-1}}$, is far in excess of the delimiter for radio-loud AGN \citep[$L_\mathrm{1.4\,GHz}>10^{23}\,\mathrm{W\,Hz^{-1}}$;][]{Best2005}, which are known to be primarily hosted by elliptical galaxies \citep[e.g.][]{Mannering2011}. These results are consistent with those of \citet{2013A&A...551A..97M} and \citet{Ruiz2017:wf} who also find that the majority of VLBI detections have early-type hosts. 

\begin{table}[tb]
	\caption{Number of multi-wavelength VLBI counterparts and sensitivities of each waveband.}             
	\label{table:VLBI_counterparts}      
	\centering 
	{\footnotesize
	\begin{tabular}{llll}
	\hline\hline
Band & $N_\mathrm{FoV}$ & $N_\mathrm{Det.}$ & $1\sigma$ Sensitivity  \\
\hline
   $K_s$ $2\micron$ & $30$ & $30$ & $0.6\microJy$ \\
   IRAC $3.6\micron$ & $28$ & $27$ & $0.06\microJy$  \\
   IRAC $4.5\micron$ & $28$ & $27$ & $0.06\microJy$ \\
   IRAC $5.6\micron$ & $28$ & $27$ & $0.66\microJy$\\
   IRAC $8.0\micron$ & $28$ & $28$ & $0.66\microJy$ \\
   WISE $3.4\micron$ & $31$ & $13$ & $80\microJy$\\
   WISE $4.6\micron$ & $31$ & $13$ & $0.11\,\mathrm{mJy}$\\
   WISE $12\micron$ & $31$ & $2$ & $1\,\mathrm{mJy}$\\
   WISE $21\micron$  & $31$ & $1$ & $6\,\mathrm{mJy}$ \\
   MIPS $24\micron$  & $24$ & $20$ & $0.5\microJy$\\
   PACS $100\micron$ &$24$ & $16$ & $0.4~(0.47)\,\mathrm{mJy}$\\
   PACS $160\micron$ &$24$ & $16$ & $0.9~(0.76)\,\mathrm{mJy}$\\
   SPIRE $250\micron$ & $31$ & $4$ & $1.3~(0.92)\,\mathrm{mJy}$ \\
   SPIRE $350\micron$ & $31$ & $3$ & $4.0~(0.77)\,\mathrm{mJy}$ \\
   SPIRE $500\micron$ & $31$ & $2$ & $6.0~(0.67)\,\mathrm{mJy}$ \\
   X-ray $0.5\mbox{-}2\,\mathrm{keV}$ & $28$ & $16$ & $1.2\times 10^{-17}\,\mathrm{erg\,cm^{-2}\,s^{-1}}$ \\
   X-ray $2\mbox{-}7\mathrm{\,keV}$ & $28$ & $15$ & $5.9\times10^{-17}\,\mathrm{erg\,cm^{-2}\,s^{-1}}$\\
   X-ray 0.5-7\,keV & 28 & 18 & $3.5\times10^{-17}\,\mathrm{erg\,cm^{-2}\,s^{-1}}$ \\
   VLA 1-2\,GHz & 31 & 31 & $2\mbox{-}10\microJybm$\\
   \hline
\end{tabular}}
\tablefoot{The $N_\mathrm{FoV}$ column corresponds to the number of VLBI sources that have coverage in a particular band, whilst the $N_\mathrm{Det.}$ corresponds to the number of VLBI counterparts detected. The number of sources in the FoV of the IRAC 5.6 and 8$\micron$ is larger than the number shown in Figure~\ref{Fig:multi-wave_coverage} due to an observation of J123656+615659 being included in the \citet{Yang_photz_2014} catalogue.}
\end{table}

\section{Classification of VLBI-detected AGN}\label{Sec:AGN_class_techniques}

With these data in hand, we can now test the various AGN classification techniques used at other wavebands in the context of the VLBI-detected sample. While no single classification technique can identify the entire AGN sample \citep[e.g. see][]{Hickox2009:ir,Mendez2012:IR,Delvecchio2017:ra}, understanding the relationships between these techniques can help us identify the systematic biases of each selection technique. Furthermore, such studies are important in deriving physical parameters that depend upon cleanly separating AGN activity from star-formation.

\subsection{Optical and ultraviolet}

The majority of optical AGN classification methods typically require spectroscopy. The standard method is to use the line ratios $\rm[O\textsc{iii}] \lambda 5007/ H\beta$ versus $\rm [N\textsc{ii}] \lambda 6584 / H\alpha$, also collectively known as the BPT diagram \citep{Baldwin1981:BPT,Kauffmann2003:bpt,Kewley2013:bpt}, to identify AGN. This has been extended to other line ratios with the most commonly used being $\rm [O\textsc{iii}]/ H\beta$ versus $\rm [S\textsc{ii}] / H\alpha$ or $\rm [O\textsc{iii}]/ H\beta$ versus $\rm [O\textsc{i}]/ H\alpha$ \citep[e.g.][]{Kewley2006:bpt,Juneau2011:bpt}.

The present analysis would benefit greatly from optical-NIR spectroscopy, but current samples are limited to small numbers of AGN in GOODS-N. For example, \citet{Coil2015;AG} identified 9 AGN in GOODS-N using data from the MOSDEF survey, none of which are detected in VLBI and only two have counterparts in the 1.5\,GHz VLA observations \citep{Owen2018;gn}. The two that are detected have integrated flux densities lower than the VLBI sensitivity limit. However, these may be detected with the completion of this VLBI survey. Of the 9 AGN stated, four are identified with optical diagnostics alone.

The OPTX-survey \citep{Trouille2008}, provides optical spectroscopic observations of the 503 X-ray sources from the main X-ray catalogue of \citet{Alexander2003}. This survey currently provides the largest available catalogue of optical spectroscopic information in GOODS-N. Of the 503 X-ray sources, 298 could be classified into the following categories. Sources without any strong emission lines, that is equivalent widths (EW) of $\mathrm{[O\textsc{ii}]< 3\,\AA}$ or $\mathrm{EW(H\alpha + N\textsc{ii}) < 10\,\AA}$, are absorbers (A). Those sources with strong Balmer lines and no broad or high-ionisation lines are classed as star-formers (SF). Sources with $\mathrm{[Ne\textsc{v}]}$ or $\mathrm{[C\textsc{iv}]}$ lines, or strong $\mathrm{[O\textsc{iii}]}$ $(\mathrm{EW([O\textsc{iii}]\lambda5007) > 3EW([H\textsc{ii}])}$ are high-excitation sources (HEG). Finally, those sources with optical lines having full-width half-maximum (FWHM) line-widths greater than $2000\,\mathrm{km\,s^{-1}}$ are broad-line AGNs (BL). In total, 17 VLBI sources have counterparts with 13 have optical classifications. Of these, only two show signs of AGN-related high excitation emission lines, with no sources showing optical broad lines. However, this sample is highly incomplete due to the X-ray selection criterion. Due to this incompleteness, and the paucity of other publicly available data, we exclude any further analysis of optical identification methods in this paper.

\subsection{Infrared}\label{SSec:Infra-red}

\begin{table}[htb]
\centering
\caption{IR AGN colour-colour classification schemes.}        
\label{table:IR_color_schemes}      
\begin{tabular}{cp{25mm}}
\hline\hline
\multicolumn{2}{c}{\textit{Spitzer} IRAC power law} \\
\multicolumn{2}{c}{\citet{alonso2006infrared,Donley2007:IR}} \\
\hline
Power law &  
$\begin{aligned}
\alpha & < -0.5~{\rm where}~ S_\nu \propto \nu^{\alpha} \\
P_\chi &< 0.1 \\
\end{aligned}$\\
\hline\hline
\multicolumn{2}{c}{\textit{Spitzer} IRAC colour-colour (flux densities)} \\
\multicolumn{2}{c}{\citet{LacyIRAC2007,donley2012identifying};}\\
\multicolumn{2}{c}{\citet{Kirkpatrick2012IR}}\\
\multicolumn{2}{c}{$x = \log_{10}\left(S_{\rm 5.8\mu m}/S_{\rm 3.6\mu m}\right),~ y = \log_{10}\left({S_{8{\rm\mu m}}}/{S_{\rm 4.5\mu m}}\right)$}\\
\hline
\citetalias{LacyIRAC2007} &
$\begin{aligned}
	x &\geq -0.1\\
	y &\geq -0.2\\
	y &\leq (0.8\times x) + 0.5\\
\end{aligned}$\\
\hline
\citetalias{donley2012identifying} & 
$\begin{aligned}
	x &\geq 0.08 \\
	y &\geq 0.15\nonumber \\
	y &\geq (1.21 \times x) - 0.27\nonumber \\
	y &\leq (1.21 \times x) + 0.27\nonumber \\
	&S_{\rm 4.5\mu m} > S_{\rm 3.6\mu m}~{\rm and}~S_{\rm 5.8\mu m} > S_{\rm 4.5\mu m}\\
	&{\rm and}~\nonumber S_{\rm 8.0\mu m} > S_{\rm 5.8\mu m}\nonumber.\\
\end{aligned}$\\\hline
\citetalias{Kirkpatrick2012IR} & 
$\begin{aligned}
x &\geq 0.08 \\
y &\geq 0.15 \\
 \end{aligned}$\\\hline\hline
 \multicolumn{2}{c}{\textit{Spitzer} IRAC colour-colour (magnitudes)}\\
  \multicolumn{2}{c}{\citet{stern2005mid}}\\
  \multicolumn{2}{c}{$x = [5.8]\vega-[8.0]\vega~,~ y = [3.6]\vega-[4.5]\vega$} \\
 \hline
 \citetalias{stern2005mid} & 
 $\begin{aligned}
 x &\geq 0.6\\
 y &\geq (0.2 \times x) + 0.18\\
 y &\geq (2.5 \times x) - 3.5\\
 \end{aligned}$\\\hline\hline
 \multicolumn{2}{c}{$K_s$-IRAC-MIPS}\\
 \multicolumn{2}{c}{\citet{Messias2012:ir,Messias2013:ir}}\\
 \multicolumn{2}{c}{$x = [K_s]\ab-[4.5]\ab,~y=[4.5]\ab-[8.0]\ab,$}\\
 \multicolumn{2}{c}{$w=[8.0]\ab - [24]\ab$}\\
 \hline
 \multirow{3}{*}{KI} &
 {For $z\leq 2.5$ only}:\newline
 $\begin{aligned} 
&x\geq0 \\ 
 &y\geq0 \\
 \end{aligned}$\\\hline
 \multirow{4}{*}{KIM} &
 {For $0 \leq z \leq 7$:} \newline 
 $\begin{aligned}
  x &> 0\\
  w &\geq(-2.9\times y)+2.8\\
  w&\geq 0.5 \\
 \end{aligned}$ \\ \hline\hline
 \multicolumn{2}{c}{WISE} \\
 \multicolumn{2}{c}{\citet{Mateos2012:irx,Jarrett2011:WISE};}\\
 \multicolumn{2}{c}{\citet{Stern2012:ir}}\\
 \multicolumn{2}{c}{$x=[4.6]\vega-[12]\vega,~y=[3.4]\vega-[4.6]\vega$} \\
 \hline
 \citetalias{Mateos2012:irx} & 
 $\begin{aligned}
 y &\geq 0.315x - 0.222 \\
 y &\leq 0.315x + 0.796 \\
 x & \geq 2.4035 - 0.315y \\ 
 \end{aligned}$\\
 \hline
\citetalias{Jarrett2011:WISE}  & $\begin{aligned} 
  x&\leq1.7\\
  y&\geq 2.2,~y\leq4.2 \\ 
  y &\geq 0.1x + 0.38 \\
  \end{aligned}$\\
  \hline
 \citetalias{Stern2012:ir}  & $\begin{aligned}y &\geq 0.8 
 \end{aligned}$ \\
 \hline
	\end{tabular}
\end{table}

Infrared classification schemes have often been used to distinguish between AGN dominated and SF dominated galaxies at high redshift. The most common schemes are the \textit{Spitzer} IRAC colour-colour schemes \citep[e.g.][]{stern2005mid,Lacy2004:ir,LacyIRAC2007,PopeIR2008,donley2012identifying,Kirkpatrick2012IR}, IRAC power-law \citep{alonso2006infrared,Donley2007:IR}, the composite $K_s$, IRAC and MIPS schemes \citep[e.g.][]{PopeIR2008,Messias2012:ir,Messias2013:ir}, and the WISE colour-colour schemes \citep{Mateos2012:irx,Stern2012:ir,Assef2013:ir}.

As briefly explained in Section~\ref{Sec: P2-Introduction}, these take advantage of the dip in the SED between the 1.6\,$\mu$m stellar emission and the longer wavelength emission from the $25\mbox{-}50\,\mathrm{K}$ cold dust heated by star-formation. However if a luminous AGN is present, this dip ceases to exist and instead a monotonic power-law SED is found in the MIR bands \citep[e.g.][]{Haas1998:ir}. This is a consequence of the UV radiation from the central radiation field of the AGN being reprocessed into the IR band \citep[e.g.][]{Pier1992:IR}. The extent of this flattening towards a power-law SED depends on the relative contribution to the MIR flux between the AGN and its host galaxy \citep[e.g. see Figure~1 from][]{donley2012identifying}. For the IRAC-only schemes, a total of 24 VLBI sources were considered while for the $K_s$+IRAC (KI) and the $K_s$+IRAC+MIPS (KIM) schemes, 24 and 20 sources were considered, respectively.

\subsubsection{IR power law galaxies}\label{SSSec:IRACpower_law}

A standard way of identifying an AGN contribution in the IR bands is the power-law selection technique \citep{alonso2006infrared,Polletta20016:IR,Donley2007:IR}. First used by \citep{alonso2006infrared}, this identifies IR power law AGN by fitting a power law to the \textit{Spitzer} IRAC bands. A source is classified as an AGN if $\alpha< -0.5$. \citet{Donley2007:IR} imposed a more stringent fitting constraint of $P_\chi \leq 0.1$, where $P_\chi$ is the probability that a fit to a power-law distribution would yield a value greater than or equal to the observed $\chi^2$. 

As explored by \citet{donley2012identifying}, and \citet{Mendez2012:IR}, it was found that the number density of power-law selected AGN does not evolve smoothly with flux as a result of the estimated uncertainties on the flux values. As the survey sensitivity improves, the respective flux density uncertainties decrease and often the $\chi^2$ values increase. This means that, within a given flux range and a fixed survey area, a shallow survey (with larger relative uncertainties) would detect more power law AGN when compared to a deeper survey (with smaller relative uncertainties). A solution is to add a 10\% uncertainty on all flux measurements \citep{donley2012identifying}.

We tested the power law selection for the 24 VLBI sources that have clear detections in all IRAC bands. Robust linear regression used the \texttt{scipy} routine \texttt{ODR} and the 10\% uncertainty modification was applied. Seven were excluded due to a poor fit ($P_\chi \geq 0.1$) and of the remaining 17, only 6 (25\%) were classified as AGN. This reveals that the power-law technique is only useful for those sources where the AGN truly dominates the AGN emission, and will miss those with a mixture of AGN and SF within the IRAC bands. In addition, the power-law technique seems to fail for those dust-poor (early-type) host galaxies where the torus emission is simply not present (see Section~\ref{SSSec:IRACcolour}).

\begin{figure*}
	\centering
	\includegraphics[width=\hsize]{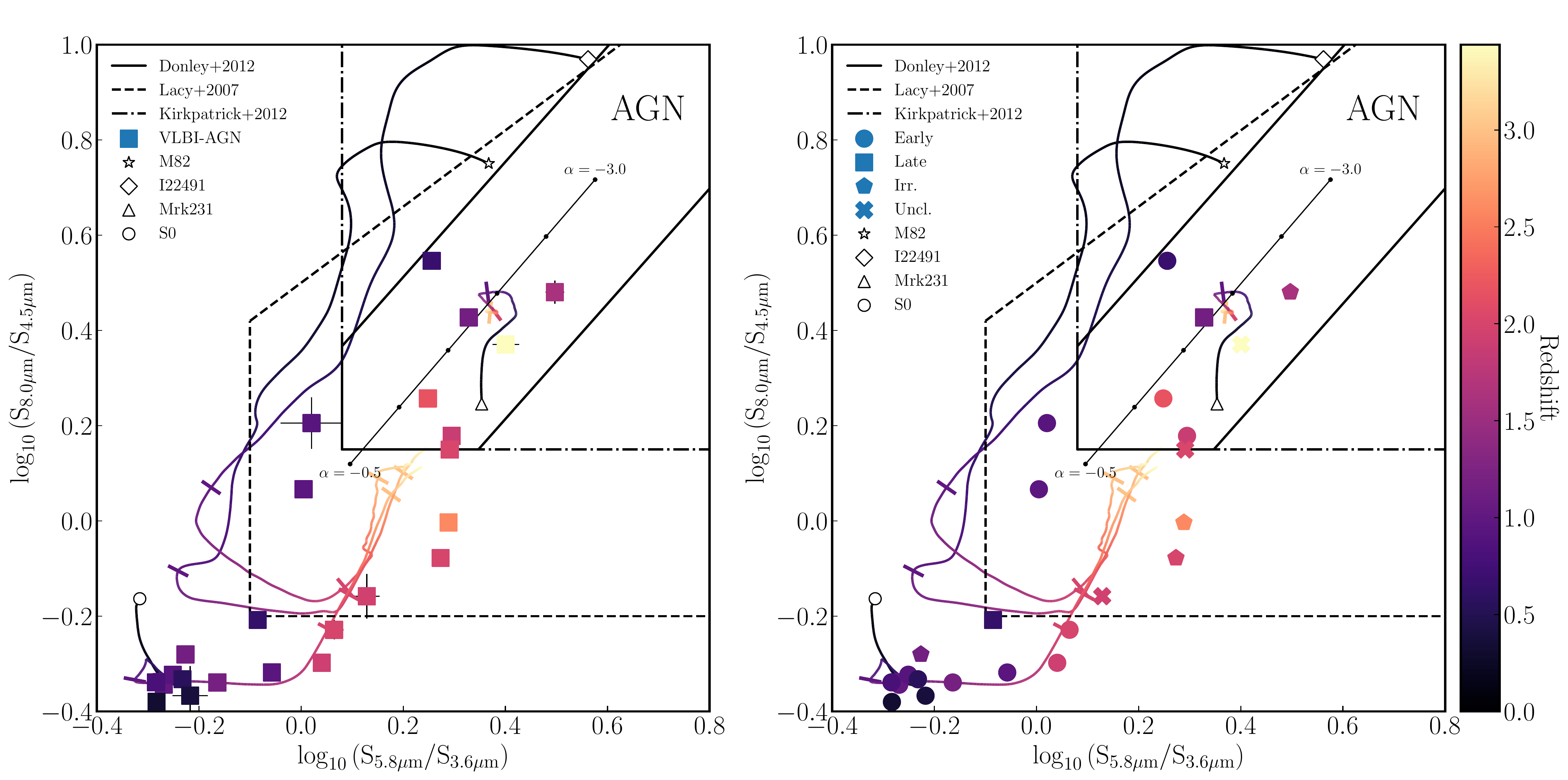}
	\caption[\textit{Spitzer} IRAC AGN selection criteria for the VLBI detected sample.]{\textit{Spitzer} IRAC AGN selection criteria for the VLBI detected sample. The dashed line marks the region where a source is classified as an AGN by \citetalias{LacyIRAC2007}, the solid line is the revised AGN selection wedge by \citetalias{donley2012identifying} and the dot-dashed line is the criterion used by \citetalias{Kirkpatrick2012IR}. For all selection methods, AGN are those within the wedges in the top-right for all selection methods The solid line with black markers corresponds to the IR power law locus of \citet{alonso2006infrared} in the range $ -3.0\leq\alpha\leq-0.5$. Overlaid are the predicted SED colours of the ULIRG IRAS22491, starburst galaxy M82, AGN Mrk231, and an S0 galaxy from the SWIRE library \citep{Polletta2007:sed} across the same redshift range as the VLBI detections ($z=0\mbox{-}3.44$). The perpendicular bars correspond to integer redshift intervals and open symbols correspond to the template MIR colours at $z=0$. \textit{Left Panel:} VLBI-selected AGN colour coded by redshift, illustrating the large fraction of VLBI selected of AGN that are missed by the IRAC selection. \textit{Right Panel:} The same VLBI-selected AGN sample instead plotted by host galaxy morphology.}\label{Fig:IRAC color color}
\end{figure*}

\subsubsection{IRAC colour-colour selection}\label{SSSec:IRACcolour}

The IR colour-colour flux ratios of $\log_{10}(S_{8{\rm\mu m}}/S_{\rm4.5\mu m})$ against $\log_{10}(S_{\rm 5.8\mu m}/S_{\rm 3.6\mu m})$, where $S_{x{\rm\mu m}}$ corresponds to the IRAC flux density at $x$ microns, has been used by many studies as an alternative to the IR power law selection technique. The \citet[][hereafter \citetalias{LacyIRAC2007}]{LacyIRAC2007} selection criterion was empirically determined using the IRAC colours of 54 quasars selected from the Sloan Digital Sky Survey Data Release 1 \citep[SDSS DR1;][]{2003AJ....126.2579S} along with MIR SED modelling based upon \textit{Infrared Space Observatory} (ISO) spectra \citep{Lacy2004:ir,LacyIRAC2007,Sajina2005}. This was reviewed by \citet[][hereafter \citetalias{donley2012identifying}]{donley2012identifying} who used \textit{XMM-Newton} X-ray observations of the COSMOS deep field, along with samples of high redshift star-forming galaxies, to calibrate and refine the IRAC criterion. This resulted in a AGN selection method that is highly complete ($\sim75\%$  of the \textit{XMM-Newton} AGN are detected) and proven, via X-ray stacking, to be efficient at selecting obscured AGN. They note though that these selection techniques cannot effectively identify low-luminosity AGN with host-dominated SEDs.

In the left panel of Figure~\ref{Fig:IRAC color color}, we show the \textit{Spitzer} IRAC colour-colour diagnostics used by \citetalias{LacyIRAC2007}, \citetalias{donley2012identifying} and \citet[][hereafter \citetalias{Kirkpatrick2012IR}]{Kirkpatrick2012IR} in the context of our VLBI sources. The \citetalias{LacyIRAC2007} wedge classifies 13/24 (54\%) VLBI objects as AGN, and the \citetalias{Kirkpatrick2012IR} and \citetalias{donley2012identifying} criteria both classify 6/25 sources (24\%) of the VLBI sources.

To investigate why these selection techniques do not detect all our VLBI sources, synthetic IRAC fluxes across the VLBI-selected sources' redshift range ($0\mbox{-}3.44$) were derived using the SEDs templates of a range of nearby galaxies obtained from the SWIRE template library \citep{Polletta2007:sed}. We track the evolution of the star-formation dominated SEDs using the starburst galaxy M82 and the ultra-luminous IR galaxy (ULIRG), IRAS 22491. These have prominent emission between $6\mbox{-}8\micron$ (rest-frame) arising from polycyclic aromatic hydrocarbons (PAH) features. These features dominate the colour-colour evolution at low redshifts, but are then redshifted out of the IRAC $3.6\micron$ band above a redshift of one. The early-type S0 (dust-poor) galaxy SED is typically dominated by photospheric emission, and so have weak PAH emission. This explains their insignificant evolution between $z=0\mbox{-}1.5$. At redshifts in excess of 2, the IRAC bands begin to sample the rest-frame 1.6$\micron$ stellar bump, and the various SEDs start to become indistinguishable. For the AGN template, we use the Markarian 231 (Mrk231) template, which resembles a power law IR SED and spirals into the power law locus at high $z$.

\begin{figure*}
	\centering
	\includegraphics[width=\hsize]{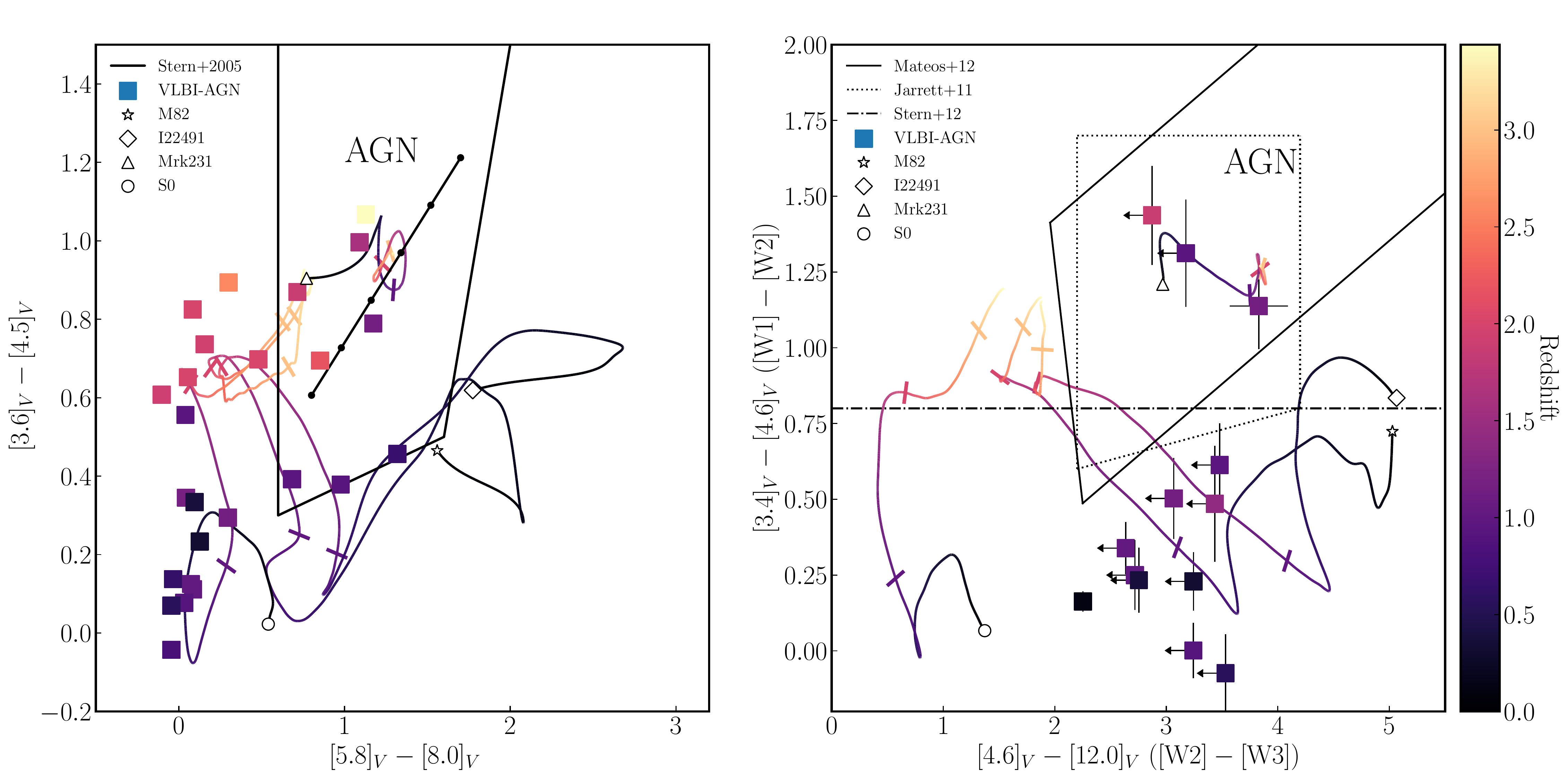}
	\caption[The \citet{stern2005mid} IRAC selection scheme and the WISE 3-band AGN identification schemes]{\textit{Left panel:} The \textit{Spitzer} IRAC AGN selection of \citep{stern2005mid}. \textit{Right panel:} the WISE 3 band colour-colour diagram. Overplotted are the AGN selection criteria of \citet{Stern2012:ir} (dash-dotted line), \citet{Mateos2012:irx} (solid line) and \citet{Jarrett2011:WISE} (dotted line)}
	\label{IRAC color color Stern}
\end{figure*}

As Figure~\ref{Fig:IRAC color color} shows, the \citetalias{donley2012identifying} wedge avoids the evolution of the starburst/ULIRG galaxy templates until redshifts around 3.5. However, the \citetalias{LacyIRAC2007} wedge suffers from significant contamination from star-forming galaxies at many redshifts and only effectively separates early-type galaxies with quiescent IR colours from those with PAH or AGN-driven IR colours. The \citetalias{Kirkpatrick2012IR} selection was designed for galaxies with redshifts in excess of 0.5 and, indeed the starburst templates are not located in this selection above this redshift. Again, this selection method can suffer from contamination at very high redshifts in excess of 3.5. On top of this, recent studies have shown that $z=2\mbox{-}3$ dust-obscured star-forming galaxies can spiral in to the AGN selection wedges \citep[e.g.][]{2019MNRAS.487.4648S,2020MNRAS.494.3828D}.

In the right panel of Figure\,\ref{Fig:IRAC color color}, we plot the host galaxy morphologies of these VLBI detected sources, overlaid onto this colour-colour space. The majority of sources follow the redshift evolution of the various templates. Those sources in the \citetalias{donley2012identifying} wedge have IR colours driven by excess AGN MIR emission, as shown by the Mrk231 AGN template. This plot also highlights that this selection technique preferentially selects only the most luminous MIR AGN. There are two early type galaxies, with $z\sim1$ and $0.0<\log_{10}(S_{8.0\micron}/S_{4.5\micron})<0.2$, which deviate from the expected early-type galaxy template. While this could be due to PAH features, both of these sources are obscured X-ray AGN, indicating that there may be a torus providing some excess MIR flux. Crucially these are not sufficiently luminous for the object to be classed as an AGN in the \citetalias{donley2012identifying} or \citetalias{Kirkpatrick2012IR} wedges. 

Another well used \textit{Spitzer} IRAC selection technique is the \citet[][hereafter \citetalias{stern2005mid}]{stern2005mid} selection criterion which uses the $[3.6]\vega - [4.5]\vega$ and $[5.8]\vega - [8.0]\vega$ colour space. This scheme was empirically defined using the IR colours of optically selected (broad and narrow line) AGN selected from the AGES survey \citep{2012ApJS..200....8K}. It was found to be remarkably good at separating normal and active galaxies with over 90\% of the broad-line AGN identified. We find it classifies a higher number of VLBI sources (8/25) compared to the \citetalias{donley2012identifying} wedge. However, as shown in the left panel of Figure~\ref{IRAC color color Stern}, this may come at a sacrifice of completeness. The SEDs illustrate possible contamination from starburst galaxies between redshifts of 1-2 whose strong PAH emission lines produce very red $[5.8]\vega-[8.0]\vega$ colours. In this colour space, the distinction between the early-type galaxies and IR AGN are apparent, with the well-known second vertical sequence (to the left of the AGN wedge) easily visible. This sequence is due to massive galaxies at $z>1.2$ \citep[e.g.][]{stern2005mid,Eisenhardt2008:ir,Papovich2008:ir}, which matches to the VLBI host morphologies as expected.

\subsubsection{WISE}\label{SSSec:WISE}

Figure~\ref{IRAC color color Stern} (right panel) shows the commonly used AGN selection criteria using WISE colours. The \citet[][hereafter \citetalias{Jarrett2011:WISE}]{Jarrett2011:WISE} and \citet[][hereafter \citetalias{Mateos2012:irx}]{Mateos2012:irx} selection criteria are severely limited due to the relatively poor sensitivity of the $12\micron$ band. As a result, only two VLBI sources can be considered, of which one is classified as an AGN. The one-colour \citet[][hereafter \citetalias{Stern2012:ir}]{Stern2012:ir} selection criteria is able to classify more sources due to its reliance only on the more sensitive W1 and W2 bands. However, as Figure~\ref{IRAC color color Stern} illustrates, this selection criteria will have some contamination from high redshift starburst galaxies. This selection criteria classifies 3/13 sources as AGN (of which 2/3 are classified by the other IR AGN selection schemes). This scheme is inherently limited compared to the IRAC selection due to the differing sensitivities between the instruments. However, it does have the significant advantage of all sky coverage, thus allowing sources outside of the IRAC coverage to be evaluated.  

Interestingly, J123726+621129, a Faranoff-Riley type I (FR-I) radio galaxy with large-scale radio lobes, is classified as an AGN using the \citetalias{Stern2012:ir} WISE selection criterion ($[3.4]\vega-[4.6]\vega = 1.31\pm0.18$), but is not classified as an AGN in any other IR selection scheme. We inspected the WISE maps and IRAC maps of this source and found no clear evidence of blending in the W1 and W2 bands that would cause this observed colour difference. The comparable IRAC colour, $[3.6]\vega-[4.5]\vega$, is 0.556 for this source, which suggests that it could have exhibited some variability between the IRAC and WISE observations.

\begin{figure*}
	\centering
	\includegraphics[width=\hsize]{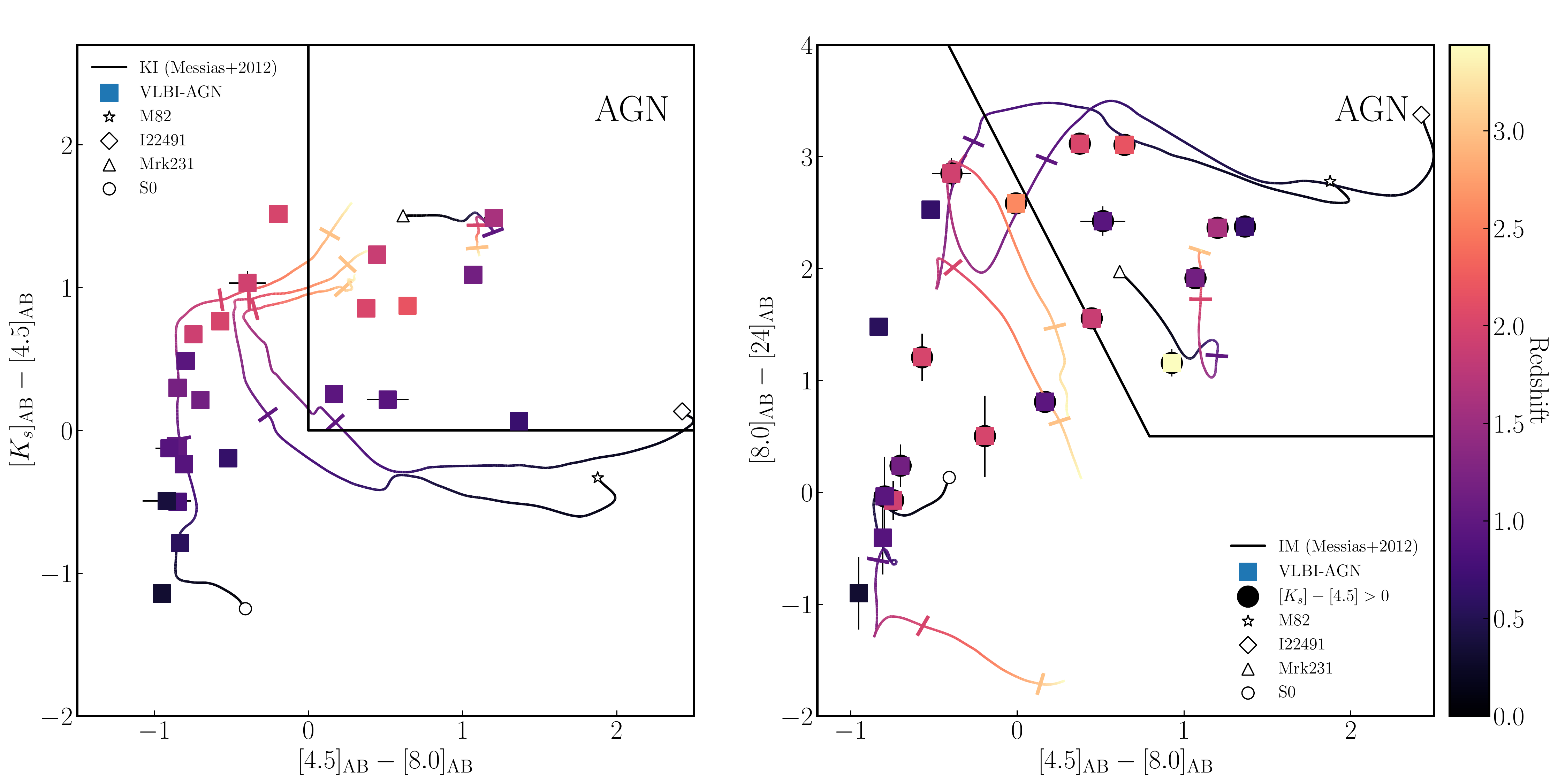}
	\caption[The KI and KIM selection schemes \citep{Messias2012:ir}.]{The KI and KIM selection schemes \citep{Messias2012:ir}. \textit{Left Panel:} The KI classification scheme suitable for IR AGN at $z\leq2.5$. Due to the redshift distribution of these VLBI sources only one is excluded from this plot. \textit{Right Panel:} The composite 3-colour KIM selection scheme plotted in the IM colour space. Sources are classed as AGN if they satisfy the K criterion ($[K]-[4.5]>0$; black circles surrounding the markers) and are located in the IM `wedge'.}
	\label{Fig:MessiasKIKIM}
\end{figure*}

\subsubsection{KI and KIM}\label{SSSec:KI_KIM}

Composite $K_s$ + IRAC + MIPS colour schemes were proposed by \citet{Messias2012:ir}. These methods are based upon a diverse range of SED templates in order to derive highly complete AGN selection techniques. These extend the wavelength coverage to different wavebands in order to overcome the shortcomings of the IRAC-only selection schemes whilst also taking into account photometric errors when deriving the selection regions. This improves the efficiency in faint source selection.

The first method, the $K_s$-IRAC (KI) criterion, is designed to select AGN at $z\leq2.5$. This redshift cut was chosen as this technique suffers from contamination above this redshift, where the stellar bump in high redshift normal galaxies can mimic a IR power-law AGN at $z>2.5$ (as noted in \citet{Messias2012:ir} and shown in Figures~\ref{Fig:IRAC color color} and \ref{Fig:MessiasKIKIM}). The advantage of this scheme is the inclusion of the $K_s$ band. This band provides a better measure of a stellar dominated waveband which can then be compared to longer wavelengths (with contributions from both AGN and stellar light). This comparison should yield a larger colour dispersion that makes it easier to separate between AGN and stellar dominated systems. It was shown to be of comparable completeness to the IRAC selection schemes (50-60\%), but less prone to non-AGN contamination ($>50\mbox{-}90\%$ successful AGN selection). For our VLBI detected sample, the KI criteria classifies 8/24 (33\%).

The second method, the $K_s$-IRAC MIPS (KIM) criterion is a 3-colour selection technique designed to select AGN hosts from redshifts from 0-7. Using an X-ray selected sample, this scheme was found to be extremely reliable ($>70\mbox{-}90\%$) at the cost of low completeness ($\sim 30\mbox{-}40\%$). The scheme has significant advantages over IRAC only schemes. Firstly, the inclusion of $[K_s]-[4.5] > 0$ selection is required to reject $z<1$ normal galaxies from the IRAC-MIPS colour space (as shown in Figure~\ref{Fig:MessiasKIKIM}). Secondly, the use of the longer wavelength MIPS-$24\micron$ mitigates the sampling of the rest-frame 1.6$\micron$ stellar bump, which causes contamination by normal galaxies at high redshifts. In the context of the VLBI sources, this selection criteria selects 8/25 (32\%) VLBI detected sources as AGN. In particular, this scheme detects the two AGN that a located near the \citetalias{donley2012identifying} wedge that may have moderate contributions from a dusty torus. 


\subsection{X-rays}

X-rays provide one of the most powerful methods of identifying AGN and currently holds the record for the highest AGN source density \citep[$\sim$25,000$\,\mathrm{deg^{-2}}$;][]{Luo2017:xray}. X-ray production in AGN originates primarily from the accretion disk where the UV photons from the accretion disk are up-scattered (inverse Compton scattering) into X-ray energies \citep[e.g.][]{TurnerMiller:xray,Gilfanov:xray}. X-ray emission can also occur in jets, and can have been detected in low-luminosity AGN where the accretion upon the central back hole is advection-dominated \citep[e.g.][and references therein]{Done2007:xray,Yuan2014}.

With regards to our VLBI selected sample, {\it Chandra} X-rays observations detect 64\% (18/28) of the sources. It is worth noting that this is a considerably higher detection fraction compared to the Very Long Baseline Array (VLBA) observations of the COSMOS field, which detect X-ray counterparts for $\sim30\%$ of the VLBI sources \citep{Ruiz2017:wf}. We believe this is due to the difference between the sensitivities of the X-ray observations. The COSMOS field has a limiting 0.5-10\,keV flux of $8.9\times 10^{-16}{\rm \,erg\,cm^{-2}\,s^{-1}}$. If we use this cut-off threshold on these GOODS-N observations, we find that 32\% (9/28) have X-ray counterparts, consistent with the COSMOS-VLBA results \citep{Ruiz2017:wf}.

Of the 18 sources detected, 14 were detected in both the soft (0.5-2\,keV) X-ray band and hard (2-7\,keV) X-ray bands, 2 were detected in only the soft X-ray band (J123716+621512 and J123701+622109), 1 was detected in only the hard X-ray band (J123715+620823) and the final source was only detected in the full band (J123641+621833). For those sources with no full band flux, \citet{xue_xray_2016} estimated the absorption-corrected X-ray luminosities by extrapolating the soft and hard-band fluxes. \citet{xue_xray_2016} provides a basic estimate of the likely source type. Here they classify a source as an AGN if it satisfies at least one of the following conditions: $L_{\mathrm{0.5\mbox{-}7keV}} \geq 3\times10^{42}\,\mathrm{erg\,s^{-1}}$, as local purely star-forming galaxies have intrinsic luminosities that are lower than this value, or an X-ray hardness with $\Gamma \leq 1$, or where $\log_{10}(f_X/f_R)>-1$, where $f_X$ and $f_R$ are the X-ray flux (in any band) and $R$-band flux respectively, or where $L_{\mathrm{0.5\mbox{-}7keV}} \gtrsim 3\times (8.9\times10^{17} L_{\mathrm{1.4\,GHz,r}})$ where $L_{\mathrm{1.4\,GHz,r}}$ is the rest-frame 1.4\,GHz monochromatic radio luminosity. Using these criteria, 16/18 X-ray detections are classified as AGN, with the remaining two (J123653+621444 and J123716+621512) categorised as `galaxies' where the origin of X-ray emission is uncertain. X-ray observations alone can only classify $57\%$ (16/28) VLBI-selected sources as definitive AGN.

As only $64\%$ of our VLBI-detected AGN are even detected in X-rays, we need to understand why are all the VLBI-detected AGN not detected in X-rays, and why are some X-ray selected AGN are not detected with VLBI. We deal with the former in Section~\ref{obscured_AGN}, while the latter is dealt with in a forthcoming paper (Radcliffe et al. in prep.). 

\begin{figure*}
	\centering
	\includegraphics[width=0.49\linewidth]{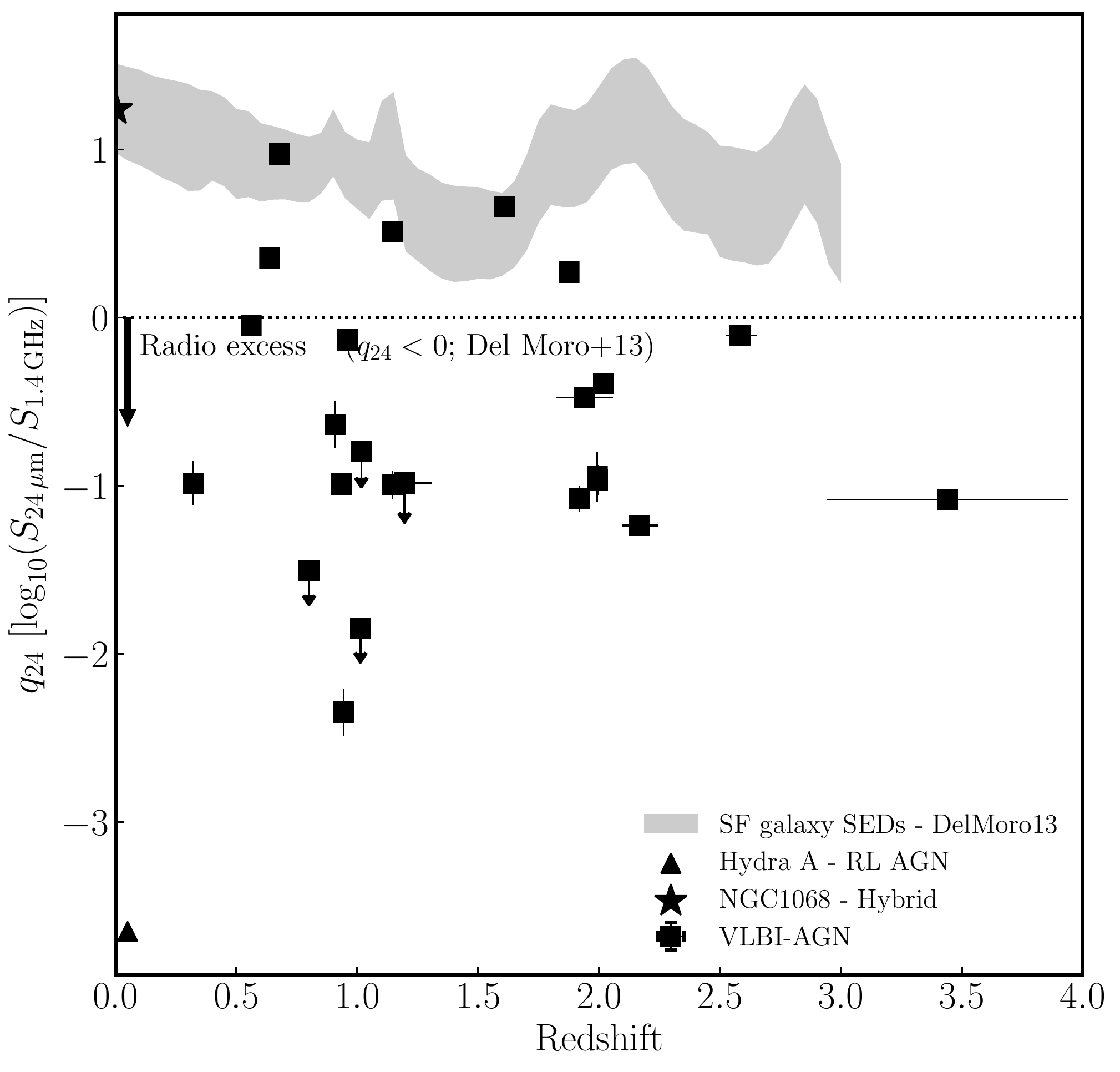}
	\includegraphics[width=0.49\linewidth]{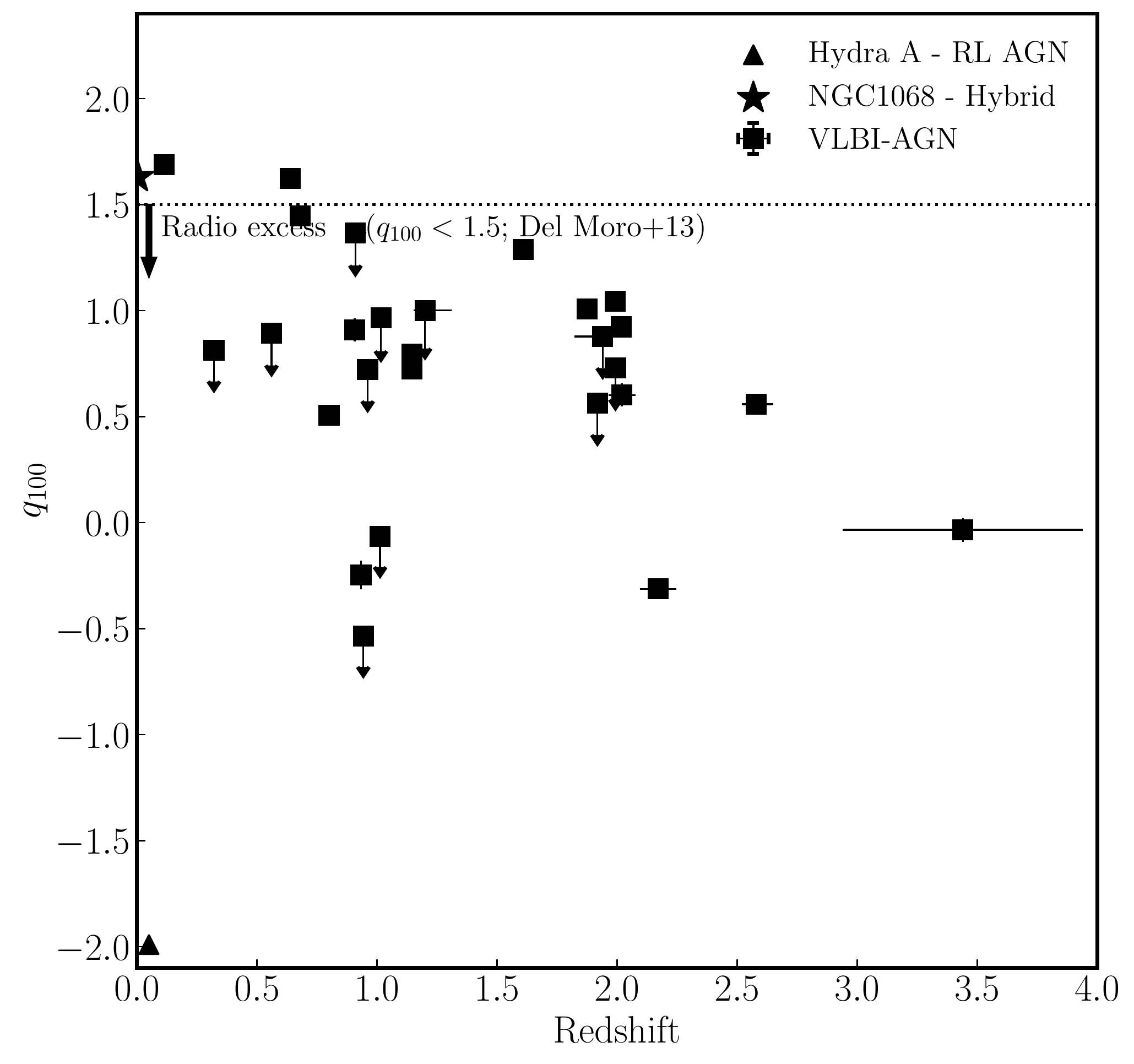}
	\caption[AGN selection using the monochromatic radio excess parameters, $q_{24}$ and $q_{100}$]{AGN selection via the radio excess parameter. The dotted lines correspond to the AGN selection criteria of \citet{Delmoro2013} and sources below this line are classified as AGN. Upper limits are denoted by the arrows. The local radio-loud AGN, Hydra-A and the hybrid system NGC\,1068 are denoted by the triangle and star markers, respectively. \textit{Left Panel:} The $q_{24}$ radio excess parameter for the VLBI selected sample. The shaded region corresponds to the $q_{24}$ evolution predicted by \citet{Delmoro2013} using a range of star-forming galaxy templates. \textit{Right Panel:} The $q_{100}$ radio excess parameter for the VLBI selected sample.}
	\label{Fig:Radio excess q_mono}
\end{figure*}

\subsection{Radio excess}\label{SSec:Radio_excess_measures}

The well known FIR-radio correlation is thought to originate from two related processes in the formation and death of massive stars ($>10\,\mathrm{M_\odot}$). The radio emission is generated via the supernovae remnants produced when these stars die, whilst the IR emission is generated by the re-processing of the UV radiation from these stars into the IR by dust. These processes are typically balanced as the starburst duration is often longer than the lifetime of these stars \citep[e.g.][]{Lacki2010}. The relation is typically parametrised with $q$, which is the ratio between the rest-frame $8\mbox{-}1000\micron$ IR flux and the 1.4\,GHz rest frame flux density. This relation is found to be invariant over four orders of magnitude and appears to hold, albeit with a mild evolution, $q\propto(1+z)^{-0.19}$, towards a redshift of 6 \citep[e.g.][]{yun2001radio,Ibar2008,Ivison2010,Magnelli2015:firrc,Delhaize2017:firrc}. It can be used as an AGN diagnostic method as those sources with an AGN present will produce excess radio emission, moving these sources away from the correlation.

Many studies have used the MIR bands, usually the \textit{Spitzer} IRAC $24\micron$, as a proxy for the FIR bolometric flux \citep[e.g.][]{Appleton24um2004, chi2013deep}, but these often have a significant contamination from SF galaxies \citep[see][]{Delmoro2013}. However, source blending and sensitivity restraints from longer wavelength instruments, such as \textit{Herschel}, limit the number of extragalactic sources where accurate bolometric FIR fluxes can be obtained. In recent times, this has been mitigated with the development of de-blending techniques \citep[e.g.][]{swinbank2014,Stanley2015:de,Thomson2017:de,Pearson2017:her,Liu2018:de}, that use prior, higher resolution, catalogues to assign \textit{Herschel} fluxes to individual sources, thus mitigating the natural \emph{Herschel} confusion limit. We adopt and compare both approaches in this section.

\subsubsection{Monochromatic radio excess - $q_{24}$ and $q_{100}$}

We define the monochromatic radio excess parameter ($q_{x}$) as,
\begin{equation}
q_{x}  = \log_{10}\left( \frac{S_{x\mathrm{\,\mu m}}}{\mathrm{Jy}}\right) - \log_{10}\left(\frac{S_{\mathrm{1.4\,GHz}}}{\mathrm{Jy}}\right), \label{Eqn:mono_radio_excess}
\end{equation}
where $S_{x\mathrm{\,\mu m}}$ is the observed flux density at $x$ microns and $S_{\rm 1.4\,GHz}$ are the observed integrated VLA flux densities at 1.4\,GHz. These flux densities were corrected for the difference in the central frequency ($\rm \sim1.51\,GHz$) assuming $\alpha=-0.7$, unless spectral index information is available from a 5.5\,GHz counterpart \citep{Guidetti:2017wt}. These were compiled for two different bands, the \textit{Spitzer} MIPS $24\micron$ and the \textit{Herschel} PACS $100\micron$. The advantage of the monochromatic radio excess measurement is that upper limits can be derived for those sources without 3$\sigma$ detections.

Figure~\ref{Fig:Radio excess q_mono} shows the $q_{24}$ parameter versus redshift. The $q_{24}$ evolutionary tracks (shown as the grey shaded region) are from \citet{Delmoro2013}. These comprise of a range of five star-forming galaxies templates from \citet{mullaney2011defining}, that have been extended to shorter wavelengths using the average starburst SED from \citet{Dale2001}. The radio emission is modelled by a power law with a spectral index of $-0.7$. Using the selection criterion of $q_{24}<0$ \citep{Donley:2005da,Delmoro2013}, a large proportion of VLBI sources are classed as AGN (79\%, 19/24). However, it is worth noting that this measure is prone to contamination by spectral features due to silicates and polycyclic aromatic hydrocarbons, along with contributions from AGN towards higher redshifts \citep[e.g.][]{Pope2006}. For example, the observed $24\micron$ emission at $z\sim2$ corresponds to a rest emission wavelength of $8\micron$, which can be influenced by power-law MIR AGN torus emission.

Longer wavelengths should be less susceptible to such contamination effects. We therefore used the deblended \textit{Herschel} PACS $100\micron$ fluxes (see Section~\ref{sect:deblending}) to calculate $q_{100}$. Only 16/27 VLBI sources had $100\micron$ counterparts. For the remaining sources, upper limits were derived. This measure is more successful, classifying 92.5\% (25/27) of the VLBI sources as AGN using the classification criteria of $q_{100} < 1.5$ \citep{Delmoro2013}. In particular, four sources that are not classed as AGN using $q_{24}$ are now classed as AGN using $q_{100}$. This illustrates precisely that $q_{24}$ can be influenced by AGN contamination. Indeed, three of these sources have AGN signatures using the MIR \citetalias{donley2012identifying} criteria (with the remaining source outside of the \textit{Spitzer} IRAC coverage). On Figure~\ref{Fig:Radio excess q_mono}, the radio excess values of Hydra A (a well known radio-loud AGN) and NGC\,1068 (an AGN and starburst hybrid system) are plotted. These show that the radio excess method can often miss hybrid systems with both emission processes present. However, VLBI provides a clean way of separating these contributions in such systems.

\subsubsection{Total infrared radio excess - $q_\mathrm{TIR}$}

\begin{figure}[tb]
    \centering
    \includegraphics[width=\linewidth]{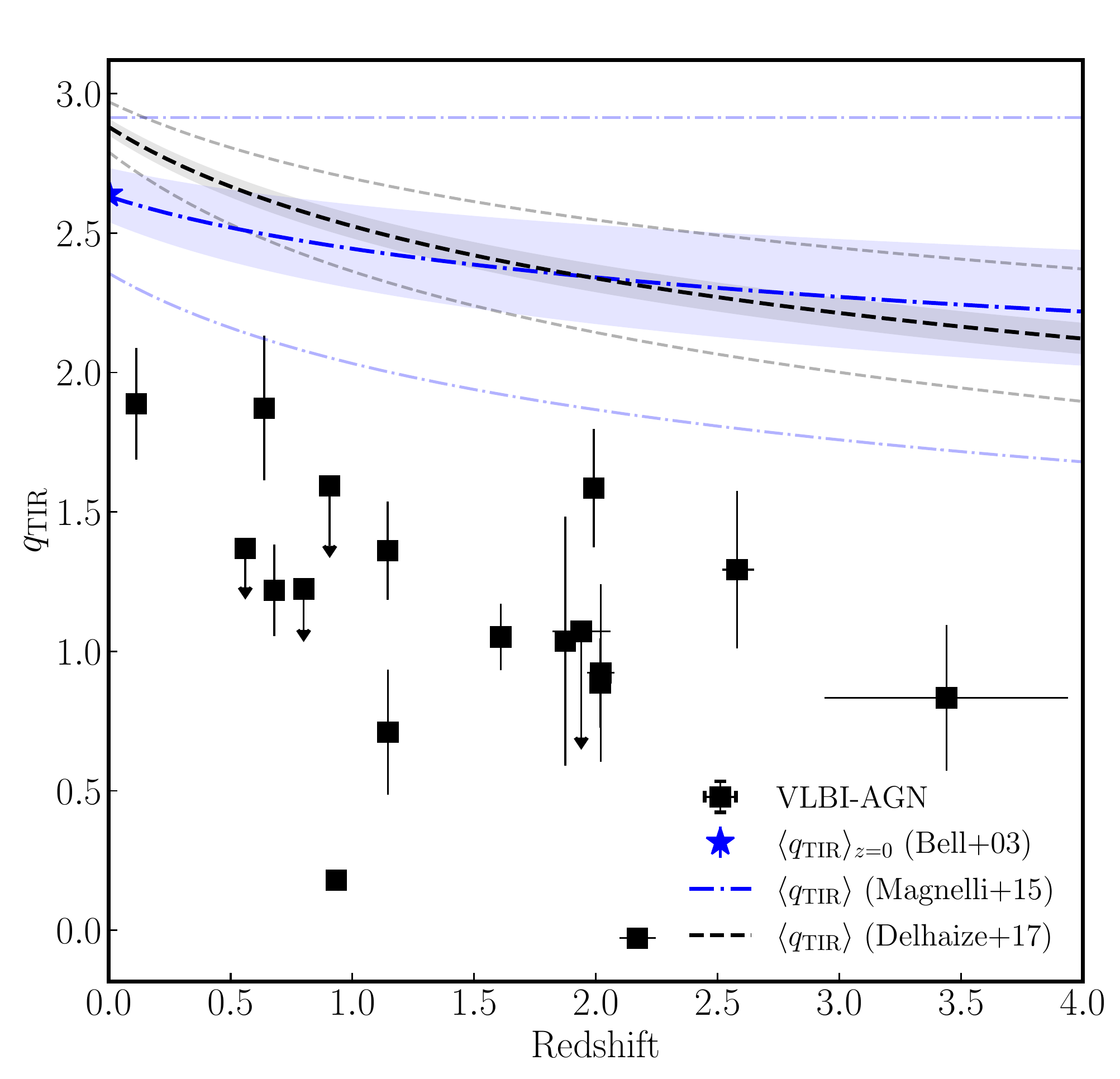}
    \caption[The total radio excess parameter $q_\mathrm{TIR}$ for those VLBI sources with FIR counterparts.]{The total radio excess parameter $q_\mathrm{TIR}$ for the subset of VLBI sources with reliable estimates of the total IR luminosity available. Over-plotted are the radio-infrared correlation evolutions from \citet{Magnelli2015:firrc} and \citet{Delhaize2017:firrc}. The filled regions correspond to the $1\sigma$ scatter, while the dotted lines correspond to the $3\sigma$ scatter. The integrated $q_\mathrm{TIR}$ parameter detects all but one VLBI source and is restricted to just a small number of sources due to their weak FIR emission.}
    \label{Fig:Radio excess q_TIR}
\end{figure}

In order to calculate the total IR radio excess measure, $q_{\rm TIR}$, we rely on total infrared luminosities ($L_{\rm TIR}$), which we measure using the deblended \textit{Herschel} PACS and SPIRE (100-500\,$\mu$m) photometry (\S\,\ref{sect:deblending}) following the method outlined in \citet{swinbank2014} and \citet{Thomson2017:de}. We begin with a composite library of AGN dominated, star-formation dominated and AGN/star-formation hybrid templates from the literature, including those of  \citet{Chary2001}, \citet{Dale2002}, \citet{Draine2007}, \citet{Reike2009:sed} and Arp220 \citep{Donley2007:IR}, as well as the Eyelash submillimetre galaxy \citep{Swinbank2010} SED. In total, this library comprises 185 distinct SED models covering a range of dust temperatures $20\,{\rm K}<T_{\rm d}<60\,{\rm K}$, where $T_{\rm d}$ is measured from the peak of the far-IR SED using the Wien approximation ($\lambda_{\rm peak}T_{\rm d}=2.987\times10^{-1}$\,cm\,K).

Next, for each EVN source in our sample we $k$-correct all template SEDs in the library to the rest frame using the published spectroscopic or photometric redshifts \citep[see][for details]{Radcliffe2018:p1}. Finally, we fit each template SED to the 100-500\,$\mu$m photometry in turn, allowing the normalisation to vary as a free parameter. We select the best-fitting template as being that which minimises $\chi^2$. For sources which are detected in three or more bands between $100$-$500\,\mu$m, we measure $L_{\rm IR}$ by integrating the best-fitting template SED using the trapezium rule, and derive the associated uncertainty by measuring the range in $L_{\rm IR}$ of all models which lie within $1\sigma$\ of the best fit\footnote{For sources with spectroscopic redshifts our SED fitting procedure has two degrees of freedom, namely the choice of template SED and the flux normalisation. For sources with only photometric redshifts, we allow the template SEDs to move within the redshift uncertainties, inducing an additional degree of freedom. Following \citet{lampton76}, we define the $1\sigma$ error bound as that which accommodates all models lying within $\Delta\chi^2=2.3$ or $\Delta\chi^2=3.5$ of the best fitting models for the cases with two and three degrees of freedom, respectively.}. 

We require detections in at least three of the five \textit{Herschel} bands to constrain the shape of the dust SED: in total, 20/31 VLBI sources had $\geq 3\sigma$ flux densities from the deblended photometry in at least three \textit{Herschel} bands, and total IR fluxes/luminosities were measured for these sources using the methods outlined above. An additional 4/31 VLBI sources had $\geq 3\sigma$ detections in two of the five deblended \textit{Herschel} maps and upper-limits in the remaining three which offers insufficient discriminatory power to choose a best-fitting SED from our library of 185 templates. In order to measure template-dependent upper-limits on $L_{\rm IR}$ for these sources, we first created a composite template by re-normalising the best-fitting SED templates for each of the 20/31 VLBI sources discussed in the preceding sentences to the same rest-frame $100\,\mu$m flux density and taking the mean of these templates as a function of wavelength, and then fit this composite SED template to the photometry/upper-limits of the 4/31 sources with detections in only two \textit{Herschel} bands. The remaining 7/31 VLBI sources have no secure detections in any \textit{Herschel} band, either because they lie outside the area of the PACS and/or SPIRE maps or because even their deblended flux densities are $<3\sigma$ upper-limits. We therefore do not attempt to fit an SED template for these sources.

The observed-frame VLA 1.5\,GHz flux densities of our sample were converted to the rest-frame 1.4\,GHz flux densities, $S_\mathrm{1.4\,GHz,r}$, assuming a spectral index of $\alpha=-0.7$, unless contradictory spectral index information was available from a 5.5\,GHz counterpart \citep{Guidetti:2017wt}. Following \citet{Ivison2010}, the bolometric radio excess parameter ($q_\mathrm{TIR}$) was then calculated using,
\begin{equation}
q_{\rm TIR} = \log _{10}\left(\frac{S_\mathrm{IR,r}}{3.75\times 10^{12}{\rm \,W\,m}^{-2}}\right) - \log_{10}\left(\frac{S_\mathrm{1.4\,GHz,r}}{{\rm W\,m}^{-2}{\rm \,Hz}^{-1}}\right).
\end{equation}
Figure~\ref{Fig:Radio excess q_TIR} shows the $q_\mathrm{TIR}$ parameter versus redshift for the VLBI sources selected here. The $q_\mathrm{TIR}$ parameter for typical star-forming galaxies from \citet{Magnelli2015:firrc} and \citet{Delhaize2017:firrc}, along with the $1\sigma$ and $3\sigma$ error bounds are over-plotted. All VLBI sources (100\%; 18/18) are classed as radio-excess sources, exceeding the $3\sigma$ scatter on both the \citet{Delhaize2017:firrc} and \citet{Magnelli2015:firrc} relations. This metric is the most complete of all the radio excess measurements, and is less susceptible to AGN contamination that can plague the monochromatic measurements \citep[e.g.][]{Delmoro2013}. However, constraints upon the fitting and the intrinsically weak MIR/FIR signatures of many objects means that $q_\mathrm{TIR}$ can only be evaluated for just over half of the sources, thus dramatically reducing its effectiveness in classifying VLBI-selected AGN.

\subsection{Radio variability}

Extra-galactic radio sources whose flux density varies is a characteristic sign of compact radio emission being present \citep[e.g.][]{Bignall2003,Koay2011}. For the same reasons as a VLBI-detection, this implies high brightness temperatures (often $>10^{12}\,\mathrm{K}$), and very small emission sizes (often $\sim \mu \mathrm{arcsec}$) which can only be attributed to AGN.

In the GOODS-N field, \citet{Radcliffe2018:var} compared the flux densities of 5 epochs of 1.5\,GHz VLA data over 22 years to investigate the variability of approximately 480 radio sources to a limiting detection threshold of $\sim 30\,\mathrm{\mu Jy\,beam^{-1}}$. In this study, a total of 10 sources were found to show significant variability. However, as this study only covered a 0.17 deg$^2$ area, and excluded those sources that are extended, only 27/31 VLBI-detected sources were included. In total, we find that 6/27 VLBI sources are classified as variables. It is worth noting that the number of VLBI sources that are variables is probably an underestimation. This is because the sparse time sampling of VLA
epochs will only detect a sub-set of any flux density variations. In addition, the variability classification metric used is rather conservative, and will not detect sources with small flux density variations ($<30\%$ difference).

Of the remaining 4 variable sources not identified by VLBI, two are below the VLBI detection threshold, while one is a 6.7$\sigma$ detection in the EVN observations and may be a supernovae \citep{Radcliffe2018:var}. The remaining variable source is undetected by the EVN observations presented here, but is detected by \citet{chi2013deep} and has changed dramatically in VLBI measured flux densities between 2004 and 2014 (from $\sim 350\,\rm\mu Jy$ to $<60\,\rm\mu Jy$).

\subsection{Radio morphologies}

A commonly used method in radio surveys is to infer the existence of an AGN based upon radio morphology alone \citep[e.g.][]{RadZoo2015:cs}. Radio-loud AGN can have large Mpc scale jets that allow them to be easily distinguished from star-formation related emission that is often confined to within the optical extent of the host galaxy. However, in deeper ($\microJy$ sensitivity) extra-galactic surveys, the number of these objects decrease rapidly as the radio population transitions from AGN to star-formation dominated regimes \citep[e.g.][]{Padovani2016}. The majority of sources are now unresolved with arcsecond resolution instruments like the VLA, that can mask AGN-related radio emission on sub-kpc scales \citep[e.g.][hereafter \citetalias{muxlow2005high}]{muxlow2005high}. Higher resolution instruments, such as $e$-MERLIN, can reveal the existence of AGN activity in these objects based upon morphology alone. In this analysis, we use the morphological analysis performed by \citetalias{muxlow2005high} who use a combination of VLA and MERLIN to identify the origin of radio emission in the GOODS-N field. This study targets 92 radio sources with integrated flux densities in excess of $40\,{\rm\mu Jy}$ based upon the 1.4\,GHz VLA observations by \citet{richards2000vla}. 

The sources were categorised into AGN/AGN candidates (AGN/AGNC), starburst/starburst candidates (SB/SBc) or unclassified objects (U). In this scheme, a source is classified as an AGN if it has a compact one or two-sided axisymmetric radio morphology that is accompanied by a flat or inverted radio spectrum (as calculated between the VLA 1.4\,GHz and 8.4\,GHz integrated flux densities). A source is classified as a starburst if it has a steep radio spectrum and is extended on sub-galactic scales. In addition, the source must have a {\it Infrared Space Observatory} (ISO) $12\,\mathrm{\mu m}$ counterpart \citep{aussel1999}. Sources with evidence of an additional embedded AGN component are classified as $\rm S^*$. Sources that do not comply with all of the characteristics are defined as AGN or starburst candidates. Finally, sources that have unclear, complex radio morphologies, which could be associated with starburst or AGN activity, are grouped into the unclassified category.

Of the 31 VLBI detected sources, 17 are included in the 92 sources considered by \citetalias{muxlow2005high}. The majority of the remaining VLBI sources were outside of the $10\arcmin\times10\arcmin$ region considered. For these 17 sources, 13/17 (76\%) are classified as AGN (9) or AGN candidates (4), whilst only one, J123642+621331, is classified as a starburst with an embedded AGN. The remaining three sources are unclassified, with these new VLBI observations confirming the existence of an AGN. The radio morphology classification scheme seems a promising way of identifying AGN but it has been constrained to small FoVs resulting in just a few sources. The extension to \citetalias{muxlow2005high}, the $e$-MERLIN Galaxy Evolution Survey ($e$-MERGE), extends this analysis to over 800 faint radio sources and so this method can be tested more robustly in the future \citep[][Wrigley et al. in prep.]{2020MNRAS.tmp.1396M}.

\begin{sidewaystable*}
	\centering
	\caption{AGN classification schemes (optical, radio and radio excess selection)}
	\begin{tabular}{l|ccc|cc|ccc}
		\hline\hline
		\multicolumn{1}{c|}{} & \multicolumn{3}{c|}{} & \multicolumn{2}{c|}{Radio} & \multicolumn{3}{c}{Radio excess}  \\
\multicolumn{1}{c|}{Source ID} & $z$ & Morphology & Opt.cl. & \citetalias{muxlow2005high} & Var. & $q_{24}$ & $q_{100}$ & $q_\mathrm{TIR}$ \\
		\multicolumn{1}{c|}{(1)} & (2) & (3) & (4) & (5) & (6) & (7) & (8) & (9) \\
\hline
J123555+620902$^{*}$ & $1.8750$ & 1 & - & - &  &                                              $0.29\pm0.04$ & $\mathbf{1.01\pm0.04}$& $\mathbf{1.04\pm0.45}$ \\
J123607+620951$^{*}$ & $0.6380$ & 2 & SF & Uncl &  &                                     $0.39\pm0.03$ & $1.62\pm0.03$ & $\mathbf{1.87\pm0.26}$ \\
J123608+621036$^{*}$ & $0.6790$ & 1 & HEG & AGN &  &                                  $0.99\pm0.03$ & $\mathbf{1.45\pm0.03}$ & $\mathbf{1.22\pm0.16}$ \\
J123618+621541$^{*}$ & $1.9930$ & 1 & - & Uncl &  &                                          $\mathbf{-0.95\pm0.09}$ & $<\mathbf{0.73}$ &  \\
J123620+620844$^{*}$ & $1.0164$ & 1 & - & AGN & \checkmark &                     $\mathbf{<-0.78}$ & $<\mathbf{0.97}$ & \\
J123621+621708$^{*}$ & $1.9920$ & 3* & - & Uncl &  &                                       $\mathbf{-0.92\pm0.15}$ & $\mathbf{1.04\pm0.04}$ & $\mathbf{1.58\pm0.21}$ \\
J123623+620654$^{*}$ & $1.94\substack{+0.12 \\ -0.12}$ & 4 & - & - & \checkmark & $\mathbf{-0.48\pm0.05}$ & $<\mathbf{0.88}$ & $<\mathbf{1.07}$ \\
J123624+621643$^{*}$ & $1.9180$ & 1 & - & AGN C & \checkmark & $\mathbf{-1.06\pm0.08}$ & $<\mathbf{0.56}$ &  \\
J123641+621833$^{*}$ & $1.1456$ & 3* & A & - &  & $\mathbf{-0.97\pm0.08}$ & $\mathbf{0.72\pm0.04}$ & $\mathbf{1.36\pm0.18}$ \\
J123642+621331$^{*}$ & $2.0180$ & 4 & - & S* &  & $\mathbf{-0.36\pm0.03}$ & $\mathbf{0.92\pm0.03}$ & $\mathbf{0.89\pm0.16}$ \\
J123644+621133$^{*}$ & $1.0128$ & 1 & A & AGN &  & $\mathbf{<-1.83}$          &      $<\mathbf{-0.06}$                  &            \\
J123646+621405$^{*}$ & $0.9610$ & 1 & SF & AGN &  & $\mathbf{-0.12\pm0.04}$ & $<\mathbf{0.72}$ &  \\
J123650+620738$^{*}$ & $1.6095$ & 3* & HEG & - &  & $0.68\pm0.02$ & $\mathbf{1.29\pm0.03}$ & $\mathbf{1.05\pm0.12}$ \\
J123653+621444$^{*}$ & $0.3208$ & 1 & A & AGN &  & $\mathbf{-0.98\pm0.13}$ & $<\mathbf{0.81}$ &  \\
J123659+621833$^{*}$ & $2.17\substack{+0.08 \\ -0.07}$ & 1 & - & - &  & $\mathbf{-1.2\pm0.04}$ & $\mathbf{-0.31\pm0.04}$ & $\mathbf{-0.03\pm0.01}$ \\
J123700+620910$^{*}$ & $2.58\substack{+0.07 \\ -0.06}$ & 3* & - & AGN C &  & $\mathbf{-0.08\pm0.04}$ & $\mathbf{0.56\pm0.04}$ & $\mathbf{1.29\pm0.28}$ \\
J123709+620838$^{*}$ & $0.9070$ & 1 & A & AGN & \checkmark & $\mathbf{-0.64\pm0.14}$             &        $<\mathbf{0.91\pm0.05}$           &     $<\mathbf{1.59}$ \\
J123714+621826$^{*}$ & $3.44\substack{+0.50 \\ -0.50}$ & 4 & - & - &  & $\mathbf{-1.06\pm0.06}$ & $\mathbf{-0.03\pm0.06}$ & $\mathbf{0.83\pm0.26}$ \\
J123715+620823$^{*}$ & $0.9335$ & 1 & - & AGN &  & $\mathbf{-0.99\pm0.06}$ & $\mathbf{-0.25\pm0.07}$ & $\mathbf{0.18\pm0.04}$ \\
J123716+621512$^{*}$ & $0.5605$ & 1* & A & AGN C &  & $\mathbf{-0.04\pm0.05}$ & $<\mathbf{0.89}$ & $\mathbf{<1.37}$ \\
J123717+621733$^{*}$ & $1.1460$ & 2 & SF & AGN C &  &           $0.54\pm0.03$ & $\mathbf{0.80\pm0.04}$ & $\mathbf{0.71\pm0.22}$ \\
J123720+620741 & $0.91\substack{+0.05 \\ -0.03}$ & 1 & - & - &  &                       -   &     $<\mathbf{1.37}$ &  \\
J123721+621130\tablefootmark{a} & $2.02\substack{+0.06 \\ -0.06}$ & 4 & SF & AGN &  &                      -&  $\mathbf{0.60\pm0.05}$ & $\mathbf{0.92\pm0.32}$ \\
J123726+621129$^{*}$ & $0.9430$ & 1 & - & AGN &  & $\mathbf{-2.31\pm0.14}$ & $<\mathbf{-0.54}$ &  \\
J123649+620439 & $0.1130$ & 1 & A & - &  &                     -      &     $1.69\pm0.01$ & $\mathbf{1.89\pm0.20}$ \\
J123701+622109$^{*}$ & $0.8001$ & 1* & A & - & \checkmark &  $\mathbf{<-1.49}$ & $\mathbf{0.51\pm0.04}$ & $<\mathbf{1.22}$ \\
J123739+620505 & $2.99\substack{+0.81 \\ -1.51}$ & 4 & - & - &  & - & - & - \\
J123751+621919$^{*}$ & $1.20\substack{+0.11 \\ -0.05}$ & 1 & - & - & \checkmark & $\mathbf{<-0.97}$ & $<\mathbf{1.00}$ &  \\
J123523+622248 & $1.42\substack{+0.10 \\ -0.11}$ & 1 & - & - &  & - & - & - \\
J123510+622202 & $2.33\substack{+0.52 \\ -0.24}$ & 4 & - & - &  & - & - & - \\
J123656+615659 & $0.39\substack{+0.05 \\ -0.04}$ & 1 & - & - &  & - & - & - \\
\hline 
& \multicolumn{2}{r}{\multirow{2}{*}{$R$}}& $17\%$ & $82\%$ & $22\%$ & $79\%$ & $92.5\%$ & $100\%$ \\
& & & $(2/12)$ & $(14/17)$ & $(6/27)$  & $(19/24)$ & $(25/27)$ & $(18/18)$ \\
\hline
\end{tabular}\label{Table:AGN_class_1}
\tablefoot{Check-marks or bold-font corresponds to a positive AGN classification. Entries with a hyphen correspond to sources which are outside the field-of-view of the bands required to classify the object or are not in the redshift range of the classification scheme. Blank entries have multi-wavelength coverage but do no have detections in all bands required. The column headers correspond to: (1) - VLBI source identifier as used in \citetalias{Radcliffe2018:p1}. (2) - Adopted redshifts from \citetalias{Radcliffe2018:p1}. Spectroscopic redshifts have no errors whilst errors on photometric redshifts correspond to 68\% confidence intervals. (3) - Optical/NIR host galaxy morphologies classified into 1. early-type / bulge dominated, 2. late-type / spiral galaxies, 3. irregular, 4. Unclassified i.e. low surface brightness or unresolved. Potential evidence of interacting systems are marked with a *. (4) - Optical classification from \citet{Trouille2008}. Optically normal galaxies (non-AGN) are classified as absorbers (A) or star-formers (SF). AGN are classified into high-excitation galaxies (HEG) or broad line AGN (BL). (5) Radio morphology classification from \citet{muxlow2005high} (AGN C - AGN candidate, S* - starburst + AGN). (6) - Radio variable sources as classified by \citet{Radcliffe2018:var}. (7) - Radio excess parameter using \textit{Spitzer} MIPS $24\micron$ emission. (8) - Radio excess parameter using \textit{Herschel} PACS $100\micron$ emission. (9) - Total radio excess parameter using rest-frame bolometric infrared luminosity and rest-frame 1.4\,GHz flux density. \tablefoottext{a}{Blending effects from a bright infrared source around 3\farcs5 from the VLBI position prevents accurate $24\micron$ fluxes to be obtained.}}
\end{sidewaystable*}

\begin{sidewaystable*}
	\centering 
	\caption{AGN classification schemes (infrared and X-ray selection)}
\begin{tabular}{l|cccccccc|c}
	\hline\hline
	&\multicolumn{8}{c|}{Infrared} & X-ray  \\
	\multicolumn{1}{c|}{Source ID} & Power law & \citetalias{LacyIRAC2007} & \citetalias{Kirkpatrick2012IR} & \citetalias{donley2012identifying} & \citetalias{stern2005mid} & KI & KIM & WISE & $L_{0.5\mbox{-}\rm 7\,keV}$  \\
	\multicolumn{1}{c|}{(1)} & (10) & (11) & (12) & (13) & (14) & (15) &(16) &(17) & (18)\\
	\hline
	J123555+620902$^{*}$ & $\mathbf{-0.96}$ & \checkmark & \checkmark & \checkmark & \checkmark & \textbf{\checkmark} & \checkmark & \checkmark & $\mathbf{3.1 \times 10^{44}\/(o)}$ \\
	J123607+620951$^{*}$ & 0.86 & $\times$ & $\times$ & $\times$ & $\times$ & $\times$ & $\times$ &  & $\mathbf{8.3 \times 10^{42}\/(o)}$ \\
	J123608+621036$^{*}$ & $-1.66$ & \checkmark & \checkmark & \checkmark & \checkmark & \textbf{\checkmark} & \checkmark &  & $\mathbf{4 \times 10^{42}\/(o)}$ \\
	J123618+621541$^{*}$ & 0.45 & $\times$ & $\times$ & $\times$ & $\times$ & $\times$ & $\times$ &  & $(\mathit{3.8 \times 10^{42}})$ \\
	J123620+620844$^{*}$ & 1.38 & $\times$ & $\times$ & $\times$ & $\times$ & $\times$ &  & $\times$ & $(\mathit{2.3 \times 10^{42}})$ \\
	J123621+621708$^{*}$ & $-0.22$ & \checkmark & $\times$ & $\times$ & $\times$ & $\times$ & $\times$ &  & $(\mathit{9.5 \times 10^{42}})$ \\
	J123623+620654$^{*}$ & 0.05 & \checkmark & $\times$ & $\times$ & $\times$ & $\times$ & $\times$ &  & $(\mathit{2.4 \times 10^{43}})$ \\
	J123624+621643$^{*}$ & 0.71 & $\times$ & $\times$ & $\times$ & $\times$ & $\times$ & $\times$ &  & $(\mathit{3.5 \times 10^{42}})$ \\
	J123641+621833$^{*}$ & 1.06 & $\times$ & $\times$ & $\times$ & $\times$ & $\times$ & $\times$ & $\times$ & $\mathbf{1.1 \times 10^{42}}$ \\
	J123642+621331$^{*}$ & $-0.76$ & \checkmark & $\times$ & $\times$ & $\times$ & \textbf{\checkmark} & \checkmark &  & $\mathbf{6 \times 10^{42}}$ \\
	J123644+621133$^{*}$ & 1.38 & $\times$ & $\times$ & $\times$ & $\times$ & $\times$ &  & $\times$ & $\mathbf{1.1 \times 10^{42}}$ \\
	J123646+621405$^{*}$ & $-0.12$ & \checkmark & $\times$ & $\times$ & \checkmark & \textbf{\checkmark} & $\times$ & $\times$ & $\mathbf{1.3 \times 10^{44}\/(o)}$ \\
	J123650+620738$^{*}$ & $\mathbf{-2.04}$ & \checkmark & \checkmark & \checkmark & \checkmark & \textbf{\checkmark} & \checkmark &  & $\mathbf{5 \times 10^{44}\/(o)}$ \\
	J123653+621444$^{*}$ & 1.42 & $\times$ & $\times$ & $\times$ & $\times$ & $\times$ & $\times$ & $\times$ & $1.3 \times 10^{41}$ \\
	J123659+621833$^{*}$ & $\mathbf{-1.02}$ & \checkmark & \checkmark & \checkmark & \checkmark & \textbf{\checkmark} & \checkmark &  & $\mathbf{5.2 \times 10^{43}\/(o)}$ \\
	J123700+620910$^{*}$ & $-0.49$ & \checkmark & $\times$ & $\times$ & $\times$ & - & $\times$ &  & $(\mathit{1.1 \times 10^{43}})$ \\
	J123709+620838$^{*}$ & 1.35 & $\times$ & $\times$ & $\times$ & $\times$ & $\times$ & $\times$ &  & $\mathbf{3.2 \times 10^{42}}$ \\
	J123714+621826$^{*}$ & $\mathbf{-1.71}$ & \checkmark & \checkmark & \checkmark & \checkmark & - & \checkmark &  & $\mathbf{2.2 \times 10^{44}}$ \\
	J123715+620823$^{*}$ & $-0.45$ & \checkmark & $\times$ & $\times$ & \checkmark & \textbf{\checkmark} & \checkmark &  & $\mathbf{1.8 \times 10^{42}\/(o)}$ \\
	J123716+621512$^{*}$ & 1.37 & $\times$ & $\times$ & $\times$ & $\times$ & $\times$ & $\times$ & $\times$ & $3.7 \times 10^{41}$ \\
	J123717+621733$^{*}$ & $\mathbf{-1.62}$ & \checkmark & \checkmark & \checkmark & \checkmark & \textbf{\checkmark} & \checkmark & \checkmark & $\mathbf{1.8 \times 10^{44}\/(o)}$ \\
	J123720+620741 &  &  &  &  &  & $\times$ &  & $\times$ & $(\mathit{3.0 \times 10^{42}})$ \\
	J123721+621130\tablefootmark{a} &  &  &  &  &  &  &  &  & $\mathbf{2.9 \times 10^{43}\/(o)}$ \\
	J123726+621129$^{*}$ & 0.86 & $\times$ & $\times$ & $\times$ & $\times$ & $\times$ & $\times$ & \checkmark & $(\mathit{1.1 \times 10^{42}})$ \\
	J123649+620439 &  &  &  &  &  &  &  & $\times$ & $\mathbf{6.1 \times 10^{40}}$ \\
	J123701+622109$^{*}$ & 1.5 & $\times$ & $\times$ & $\times$ & $\times$ & $\times$ &  &  & $\mathbf{1.5 \times 10^{42}}$ \\
	J123739+620505 &  &  &  &  &  & - &  &  & $(\mathit{1.6 \times 10^{44}})$ \\
	J123751+621919$^{*}$ & 1.13 & $\times$ & $\times$ & $\times$ & $\times$ & $\times$ &  &  & $(\mathit{6.0 \times 10^{42}})$ \\
	J123523+622248 &  &  &  &  &  &  &  & $\times$ & - \\
	J123510+622202 &  &  &  &  &  &  &  &  & - \\
	J123656+615659 & 1.2 & $\times$ & $\times$ & $\times$ & $\times$ & $\times$ &  & $\times$ & - \\
	\hline
	\multicolumn{1}{r|}{\multirow{2}{*}{$R$}} & $20\%$ & $48\%$ & $24\%$ & $24\%$ & $33\%$ &  $33\%$ & $40\%$ & $30\%$ & $59\%$  \\
	& $(5/25)$ & $(12/25)$ & $(6/25)$ & $(6/25)$ & $(8/24)$ & $(8/24)$ & $(8/20)$ & $(4/13)$ & $(16/28)$  \\
	\hline 
\end{tabular}\label{Table:AGN_class_2}
\tablefoot{Check-marks or bold-face entries correspond to a positive AGN classification. Entries with a hyphen correspond to sources which are outside the field-of-view of the bands required to classify the object or are not in the redshift range of the classification scheme. Blank entries have multi-wavelength coverage but do no have detections in all bands required. The column headers correspond to: (10) - Infrared power law classification \citep{alonso2006infrared,Donley2007:IR}. (11) - \citet{LacyIRAC2007}. (12) - \citet{Kirkpatrick2012IR}. (13) - \citet{donley2012identifying}. (14) - \citet{Messias2012:ir,Messias2013:ir} $K_s$+IRAC. (15) - \citet{Messias2012:ir,Messias2013:ir} $K_s$+IRAC+MIPS (16) - \citet{Stern2012:ir} WISE classification. (17) - 0.5-7\,keV \textit{Chandra} X-ray luminosities in $\mathrm{erg\,s^{-1}}$. Those sources denoted with an (o) are classified as obscured by \citet{xue_xray_2016} (i.e. $\Gamma < 1$) while bracketed entries correspond to X-ray upper limits.}
\end{sidewaystable*}

\subsection{Summary}

For us to compare all AGN classification techniques, we are going define the AGN-classification metric ($R$) for our VLBI sample as,

\begin{equation}
    R = \frac{N_\mathrm{AGN}}{N}~[\%],
\end{equation}

\noindent where $N$ are the number of VLBI-classified AGN that can be considered for the AGN classification and $N_\mathrm{AGN}$ are the number of positive AGN classifications. In Tables \ref{Table:AGN_class_1} and \ref{Table:AGN_class_2}, we present a summary of all of the AGN classification techniques discussed in the previous section. We find that the radio variability and IR power law techniques are the least effective with $R\sim 20\%$. This is followed closely by the other MIR selection techniques that have $R$ between 20 and 40\%. The \citetalias{LacyIRAC2007} has the highest $R$ but, as explained in \citetalias{donley2012identifying} and Section \ref{SSSec:IRACcolour}, is most likely contaminated by significant star-formation activity. The KIM metric provides the highest IR reliability without significant contamination ($R\sim40\%$). However, this method requires detections in 5 different bands in order to be evaluated which results in fewer sources being considered.

The 2\,Ms \textit{Chandra} X-ray AGN classification has the next best reliability with $R\sim59\%$. While X-ray emission is widely assumed to be a near universal property from AGNs \citep{Brandt2015} some VLBI sources remain undetected. We discuss the missing X-ray AGN in Section~\ref{obscured_AGN}. 

As expected, the radio-based AGN classification techniques are the most reliable. The \citetalias{muxlow2005high} radio morphological classification performs well with an $R\sim82\%$. However, it is worth noting that only 17 VLBI sources are contained in this study. The upcoming $e$-MERGE survey will extend this sample to the entire radio population of the GOODS-N field.

The most reliable metric studied is radio excess. The monochromatic radio excess measures increases in reliability towards longer wavelengths, with $R\sim79\%$ and $R\sim92.5\%$ for $q_{24}$ and $q_{100}$, respectively. This reflects the decreasing contamination effects of the single band IR data by AGN or redshifted PAH emission. The $q_\mathrm{TIR}$ is the most reliable, classifying all sources as AGN. However, this is severely limited due to constraints upon the number of bands required to evaluate this metric, which limits the total number of sources to 18 in this study. To conclude, this analysis clearly shows that {\it no} single AGN classification technique identifies a VLBI-selected AGN sample.

\section{The nature of the VLBI-selected population}\label{Sec:AGN_overlap}

While in the previous section we identified the reliability of each classification scheme individually, there are many biases to be addressed before we can infer about the relationships between the various classification techniques, and what they imply about the nature of the underlying AGN. The first bias is due to the varying multi-wavelength coverage and depth across the GOODS-N field (see Figure\,\ref{Fig:multi-wave_coverage}), which results in each classification technique probing a different number of VLBI-sources. To mitigate this, we select a sub-sample of 24 VLBI sources that are located within the \textit{Spitzer} $5.8\micron$ and $8.0\micron$ area of GOODS-N. This area is completely covered by the 1.5\,GHz VLA, \textit{Spitzer} IRAC and MIPS, \textit{Herschel} PACS and SPIRE, and the \textit{Chandra} observations. This subset of VLBI sources are highlighted with an asterisk in Tables \ref{Table:AGN_class_1} and \ref{Table:AGN_class_2}. We note that J123721+621130 is excluded due to a nearby bright IR source $3\farcs5$ away which prevents any reliable $24\micron$ fluxes from be obtained.

\begin{figure*}
    \centering
    \includegraphics[width=0.49\linewidth]{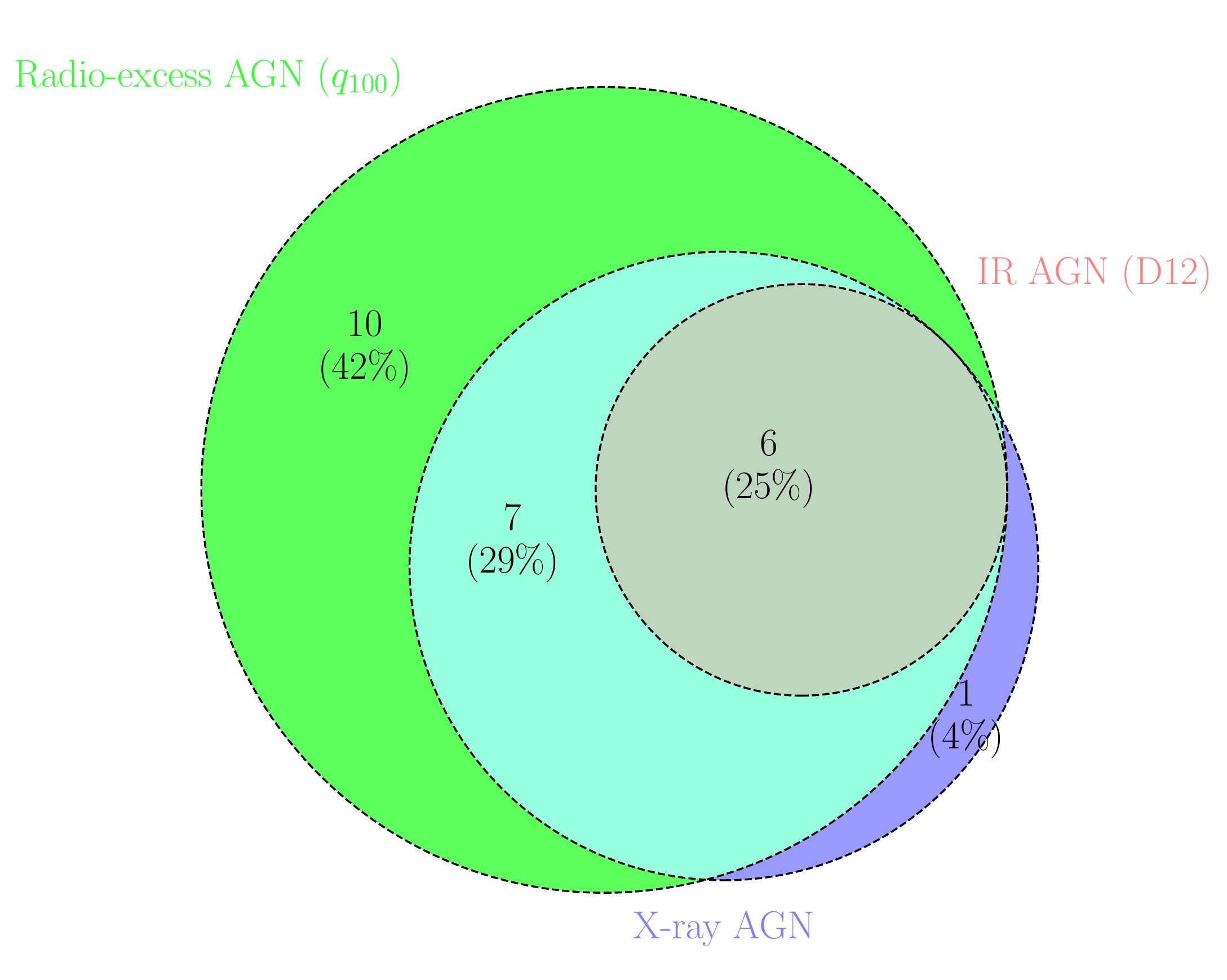}
    \includegraphics[width=0.49\linewidth]{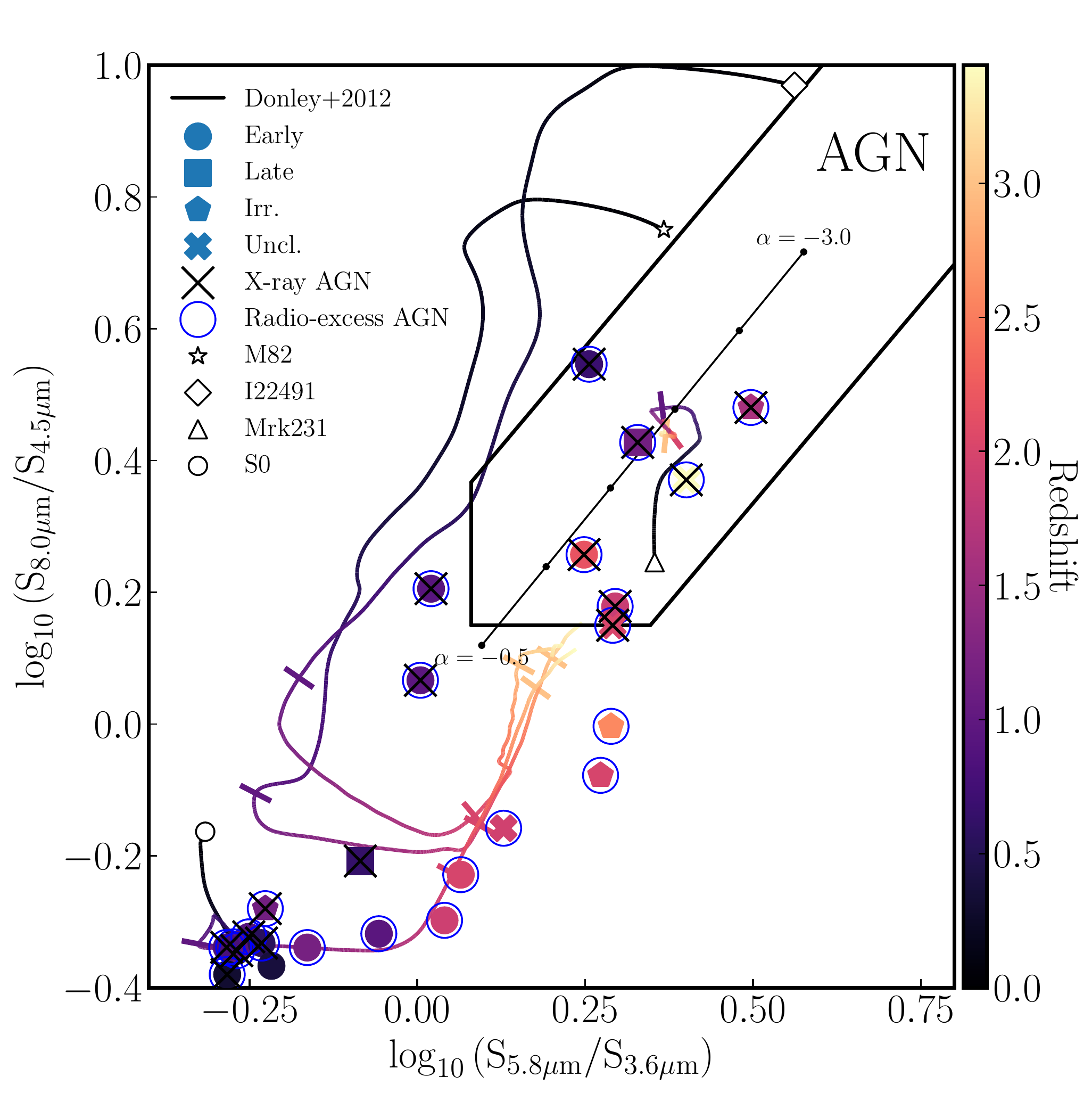}
    \caption{{\it Left Panel:} A Venn diagram showing the breakdown of the 24 VLBI-selected AGN which can be evaluated using the \citetalias{donley2012identifying} IR wedge, X-ray, and $q_{100}$ radio excess AGN classification schemes. The combination of radio excess, IR and X-ray AGN classifies all VLBI-selected AGN but crucially no one classification technique can classify all VLBI-selected AGN. {\it Right Panel:} The same 24 VLBI selected AGN in the IRAC colour-colour space with the \citetalias{donley2012identifying} wedge overlaid. All VLBI AGN in the wedge show radio-excess and X-ray signatures.}
    \label{Fig:VLBI_venn}
\end{figure*}

In Figure~\ref{Fig:VLBI_venn} (left panel), we illustrate the breakdown of the VLBI sample into the three main classification schemes (namely IR, radio excess and X-rays). We use the $q_{100}$ metric because this can be evaluated for all 24 sources and was shown to be less susceptible to contamination from AGN. The \citetalias{donley2012identifying} IR classification is used due to its proven reliability for selecting IR-AGN.

The key point from Figure~\ref{Fig:VLBI_venn} is that {\it all} VLBI sources are classified as AGN using a combination of X-ray, radio excess and IR measures. This independently verifies similar approaches used in other deep fields \citep[e.g.][]{Hickox2009:ir,Mendez2012:IR,Delvecchio2017:ra}. Indeed, \citet{Delvecchio2017:ra} used a combination of X-ray, radio excess and SED fitting to split the VLA-COSMOS 3\,GHz survey sample into AGN and star-forming galaxies, and it was found that these techniques could classify $91\%$ of the VLBA detections correctly as AGN \citep{Ruiz2017:wf}. In the following sub-sections, we shall explore a few different outcomes and questions arising from these analyses.

\subsection{Radiatively efficient versus inefficient AGN populations}

The various positive AGN classification techniques of the VLBI-selected sources in different bands can be used to infer about the nature of the VLBI-selected AGN. Observations of the local Universe have revealed that there is dichotomy in nuclear activity related to the accretion rate. AGN that are accreting efficiently (>0.01$\dot{M}_{\rm Edd}$) are collectively termed radiatively-efficient (RE-AGN) or `quasar-mode' AGN, whilst inefficient accretion (<0.01$\dot{M}_{\rm Edd}$) are known as radiatively-inefficient or `radio-mode' AGN (RI-AGN). For a review, see \citet{HeckmanBest2014:agn} and \citet{2020NewAR..8801539H}. RE-AGN are powerful and emit across multiple wavebands (from MIR to X-rays). These objects typically exhibit high excitation emission lines and have optically thick, but geometrically thin accretion disks where the gas can radiate efficiently due to the high gas density \citep{Shakura1973}. They often exhibit MIR AGN activity due to the presence of a torus with high column densities. The accretion onto the central black hole is typically from cold gas via secular processes. Most importantly, these objects are typically radio-faint, but a small proportion emit powerful relativistic radio jets. These represent the true `monsters', such as the 3C radio galaxy and quasar populations.

In contrast, RI-AGN typically do not emit across the multi-wavelength spectrum. They are typically identified by their radio emission, which manifests as jet-like radio morphologies or an excess of radio emission above that which is expected from star-formation related activity alone \citep{Hardcastle2007}. These objects are often associated with low excitation emission lines and are thought to accrete only hot gas from the galaxy halo \citep[e.g.][]{Hine1979,Laing1994,MullerSanchez2011}. The hosts of these AGN are massive and passive red sequence galaxies that are located in dense environments. AGN feedback is thought to maintain this status-quo, where episodic mechanical feedback from AGN jets transfer the energy into the surroundings, keeping halo gas temperatures high and inhibiting star-formation \citep[e.g.][]{Croton2006}.

While we cannot get a direct measurement of the accretion rate of the central black hole, we can infer whether these VLBI sources are RE- or RI-AGN using the various AGN classifiers. We follow a similar methodology of \citet{Guidetti:2017wt} who used IR, X-ray and radio-excess diagnostics to separate a 5.5\,GHz VLA sample into RE-AGN, RI-AGN and star-formation/hybrid systems. They define RE-AGN as those radio sources identified as AGN by IR diagnostics and have a 0.5-7\,keV X-ray luminosity in excess of $10^{42}\,\mathrm{erg\,s^{-1}}$. In contrast, RI-AGN are identified as those radio sources having MIR colours typical of red and passive galaxies, or those showing a radio-excess ($q_{100}<1.5$). For all other sources that do not fit into the RE-/RI-AGN classification, we classify them as undetermined instead. This is simply because we know an AGN is present but are unable to determine whether a dusty torus, typical of RE-AGN, is present.

Using these criteria we determine that 25\% (6/24) of the VLBI-selected sample are classified as RE-AGN, while 42\% (10/24) are classed as RI-AGN with the remaining undetermined. The preference in selecting RI-AGN is entirely expected because the surface density of radio-loud RE-AGN is much lower than of RI-AGN. However, even though we sample radio sources down to luminosities $10^{-8}$ of the most powerful 3C sources, we still obtain a reasonable number of efficiently accreting AGN. 

As an aside, we stress that about half of the X-ray selected RE-AGN remain undetectable at radio wavelengths, even at the currently highest achievable sensitivity \citep{2017ApJ...835...95B}. Stacking techniques have in the meantime allowed us to begin to understand their nature -- to be reported elsewhere.

\subsection{X-ray undetected sources}\label{obscured_AGN}

There is a population of Compton-thick (CT) AGN with intrinsic column densities in excess of  $>10^{24}\,\mathrm{cm^{-2}}$ that are thought to contribute towards 10-25\% of the total cosmic X-ray background \citep[e.g.][]{GilliX_ray2007,Ananna2019}. The general consensus is that the obscuration is caused by a dusty torus surrounding the central SMBH and its accretion disk. In these objects, the fraction of emission detected decreases at soft X-ray energies ($\leq 10\,\mathrm{keV}$). At these lower energies only the emission that is scattered rather than absorbed is detectable \citep{Ricci2015,Koss2016}. In principle, some of these sources must be detectable by these radio observations. However, it is worth stating that only some sources will have radio jets detectable by VLBI, so some CT-AGN will be missed. 

Almost 36\% of our VLBI sources remain undetected in the $0.5\mbox{-}7\,\mathrm{keV}$ \textit{Chandra} observations. Of those sources with X-ray counterparts, 53\% have hard X-ray spectra (an effective photon index, $\Gamma<1$), indicative of a significant level of obscuration. The median photon index of VLBI sources is 0.87, which is lower than the median photon index of the whole X-ray sample which is 0.97.

In Figure\,\ref{Fig:VLBI_venn} (right panel), we show the classifications of the 24 VLBI-detected AGN within the IRAC colour-colour space. As stated earlier, there is a cluster of sources with passive IRAC colours towards the bottom left ($\log_{10}(S_{8.0\micron}/S_{4.5\micron}) < -0.2 \wedge \log_{10}(S_{5.8\micron}/S_{3.6\micron}) < -0.1$). These sources are located at lower redshifts ($z\sim 0.8$) and have X-ray luminosities in the range of $10^{41}\mbox{-}10^{42}\ergs$. In contrast, the X-ray undetected VLBI sources have higher average redshifts ($z\sim 1.41$), but similar host morphologies. If we take those X-ray detected galaxies with passive IRAC colours, and redshift their X-ray luminosities to $z=1.41$, we find that a third of these sources would now be undetected in the $\mathrm{2\,Ms}$ \textit{Chandra} exposure. This suggests that the cause is either due to sensitivity limitations or intrinsic obscuration. 

In an attempt for clarity, we used the MIR-X-ray correlation for AGN to predict the expected X-ray luminosity of the X-ray undetected VLBI sources. We used the empirical relation from \citet{Mateos2015} which relates the rest-frame $6\,\micron$ luminosity ($L_\mathrm{6\micron}$) to the 2-10\,keV X-ray luminosity ($L_\mathrm{2-10\,\mathrm{keV}}$) by, 
\begin{equation}
	\left(\frac{L_\mathrm{6\micron}}{10^{44}}\right) = \gamma\left(\frac{L_\mathrm{2-10\,\mathrm{keV}}}{10^{44}}\right)^\beta \label{eqn:MIR_XRAY}
\end{equation}
where $\gamma = 2.003\substack{+0.120\\-0.113}$ and $\beta = 0.986\substack{+0.03\\-0.03}$. The $6\micron$ flux densities were estimated using a linear fit between the $24\micron$ and $8\micron$ flux densities and converted into a monochromatic luminosity. Using Eqn.~\ref{eqn:MIR_XRAY}, expected $2\mbox{-}10\,\mathrm{keV}$ X-ray luminosities were calculated for each undetected source and converted to 0.5-7\,keV luminosities by dividing by a factor of 0.721 \citep[see Section 3.3.4 of][]{xue_xray_2016}. 

We find that all of the undetected sources should be detectable if this relation holds (with the expected X-ray luminosity between 3-50 times the sensitivity limit). While this could be due to obscuration, the most likely effect is due to stellar contamination which boosts the IR luminosities and the subsequent expected X-ray luminosities. Indeed, for the X-ray detected sources, we find that most ($\sim 70\%$) of the expected X-ray luminosities are in excess of the measured luminosities but by a much smaller extent (approximately 5 times the measured luminosities). This suggests that the $6\micron$ band also has additional contributions from star-formation that causes the expected luminosities to exceed that of the measured luminosities. To disentangle the two, and to understand the reason for the missing X-ray sources, more sophisticated SED fitting routines that can decompose the AGN and star-forming contributions across the IR spectrum would be needed. However, such a study is outside of the scope of this paper.  

Instead, to find the reason for the X-ray non-detections, we performed X-ray stacking with the final goal of determining the X-ray hardness ratio. A high hardness ratio would be indicative of significant obscuration implying that the Compton-thick conclusion is most viable. For this analysis, we followed the steps outlined in \citet{2018MNRAS.474.4528V} which are summarised in the following paragraphs. 

\begin{figure*}
	\centering
	\includegraphics[width=\linewidth]{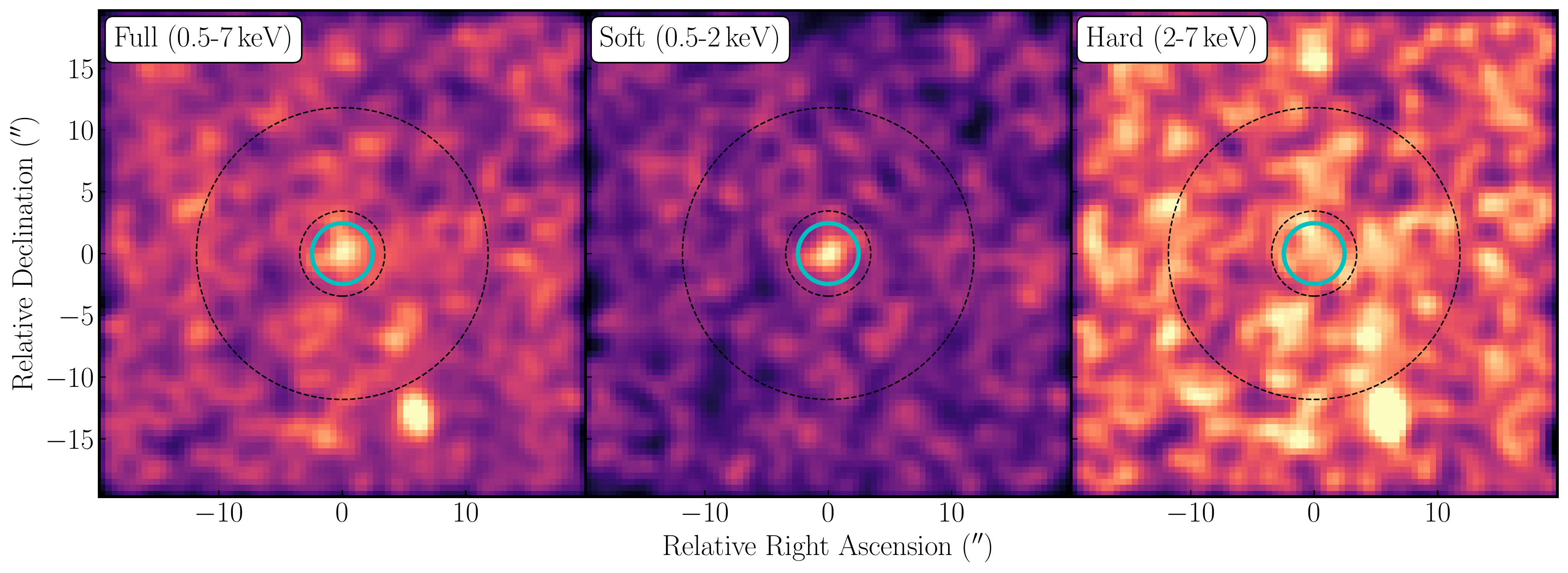}
	\caption{Stacked and smoothed X-ray images of the ten X-ray non-detections in the full (0.5-7\,keV; left), soft (0.5-2\,keV; centre) and hard (2-7\,keV; right). Each panel is approximately $40\arcsec\times40\arcsec$ in size. The blue solid circle with radius of 5\,pixels ($\sim 2\farcs5$) and dashed annulus represents the source and background extraction regions.}
	\label{Fig:X_ray_stack}
\end{figure*}

Cut-outs centred on the ten X-ray non-detections were generated for each of the three {\it Chandra} X-ray bands, soft (0.5-2\,keV), hard (2-7\,keV) and full bands (0.5-7\,keV). These cut-outs are $80\times 80$ pixels in size corresponding to approximately $40\arcsec \times 40\arcsec$. These were then summed and circular extraction regions that are $\sim5$ pixels in radius are used.These are assumed to contain all of the source counts. The background levels were derived in annuli outside of the source extraction regions. The stacks and extraction regions are shown in Figure~\ref{Fig:X_ray_stack}. Following \citet{weisskopf2007}, we assessed the detection significance using the binomial no-source probability, $P_B$, which is described by,
\begin{equation}
	P_B(X\geq S) = \sum_{X=S}^{N}\frac{N!}{X!(N-X)!}p^{X}(1-p)^{N-X}
\end{equation}
where $B$ is the total number of counts in the source region, $S$ is the total number of counts in the source region and $N = S+B$ is the total number of counts. $p=1/(1+\epsilon)$ where $\epsilon$ is the ratio of background to source region areas. The value of $(1-P_B)$ gives the significance of the detection of a source. We follow \citet{2018MNRAS.474.4528V} and set a detection threshold of 99.95\% ($\sim3.5\sigma$ for a Gaussian equivalent) which corresponds to a $P_B$ of $5\times10^{-4}$. 

Out of the three stacks, we confirmed a detection in the 0.5-2\,keV (soft) band only, with a corresponding $P_B = 2.16\times10^{-4}$. There is a tentative detection in the 0.5-7\,keV (full-band) stack with a $P_B = 5.12\times10^{-3}$ or $2.9\sigma$ in the Gaussian approximation. However, there is no statistically significant detection in the 2-7\,keV or hard X-ray band ($P_B = 0.32$ or $1.2\sigma$).

To assess the typical X-ray spectrum of these objects, we performed a hardness ratio analysis of the stacked X-ray. Using the 0.5-2\,keV and 2-7\,keV stacked signals as the soft and hard source counts, respectively, the hardness ratio is calculated using,
\begin{equation}
	\mathrm{HR} = \frac{\mathcal{H} - \mathcal{S}}{\mathcal{H} + \mathcal{S}},
\end{equation}
where $\mathcal{S}$ and $\mathcal{H}$ are the net (i.e. background-subtracted) counts in the soft and hard bands. The \texttt{BEHR} code \citep{2006ApJ...652..610P} was used to calculate this, because it accounts for the Poisson nature of both source and background counts (including non-detections). This resulted in $\mathrm{HR} = -0.66 \substack{+0.13 \\ -0.34}$. This extremely soft hardness-ratio is suggestive of a distinct lack of obscuration in these sources. This indicates that the majority of the X-ray undetected sources are most likely due to the lack of sensitivity in the X-ray band rather than being Compton-thick objects. 

\subsection{The underlying total AGN population and the role of VLBI}

In order to see where VLBI fits into the entire AGN classification scheme, we also need to reverse the analysis of the previous sections and identify the performance of VLBI in terms of the various AGN classification schemes. Due to the larger number of counterparts, we use the $q_{24}$ radio excess metric, utilising the 2.2$\microJybm$ 1.5\,GHz VLA catalogue of \citet{Owen2018;gn}. This analysis is restricted to the same region outlined in the previous subsection. For the IR-AGN, we shall use the \citetalias{donley2012identifying} metric and X-ray observations provided by the \textit{Chandra} 2\,Ms catalogue \citep{xue_xray_2016}.

In particular, the ultra-deep IRAC coverage in this area $(1\sigma_{3.6\micron} \sim 0.02\microJy)$ produces a total of 195 \citetalias{donley2012identifying} selected MIR-AGN and there are 307 X-ray selected AGN. The radio-excess measurement, $q_{24}$, reveals a total of 85 AGN.

In total, there are 507 unique AGN candidates. Around 11\% of these both X-ray and IR-selected AGN and around 16.7\% are radio-excess AGN. However, VLBI-selected AGN contribute to just 5\% (24/507) of the total AGN content. The lack of overlap between radio excess AGN, and the X-ray and IR AGN (approximately 5\% of the sample) reinforces that compact radio emission from AGN is not apparent in a large majority of AGN \citep[e.g.][]{Delvecchio2017:ra,2020ApJ...903..139A}. 

As shown in the previous section, VLBI traces mostly radio-excess AGN. However, VLBI only detects around 22\% (19/85) of the $q_{24}$ selected radio-excess AGN. The main reason why the number of VLBI sources do not match to the radio-excess is because the VLBI observations are inherently biased. The first bias due to the mismatch in sensitivity between the VLA observations ($1\sigma \sim 2\microJybm$) and the VLBI observations ($1\sigma \sim 9\microJybm$). As a result, the VLA observations classify a total of 85 sources as radio excess, while only 22\% of these are VLBI-detected AGN. If we match the point source detection thresholds of the VLA and VLBI observations ($\sim 60\microJybm$), the percentage of radio excess sources detected by VLBI increases to 61\%. Crucially though, VLBI does not detect them all.

This brings us to the second bias namely the spatial filtering effect of VLBI. The sparse nature of VLBI arrays make them insensitive to large scale structures, for example AGN jets, which all can contribute to the excess radio emission. It has been found that at $\microJy$ flux densities the fraction of VLBI to VLA flux is around 0.6 \citep{Ruiz2017:wf,Radcliffe2018:p1}. For the VLBI survey presented here, which has a detection threshold of $60\microJybm$, we effectively sample the VLA-detected radio population with peak brightnesses in excess of 100$\microJybm$. Even with a 100$\microJybm$ detection threshold, these VLBI observations only detect 19/25 (76\%) of the VLA radio excess sources. This is likely due to the large scatter (2.5\,dex) in the VLBI to VLA flux ratio \citep{Ruiz2017:wf}.

To alleviate these discrepancies for future surveys, and bring VLBI to be on par with radio excess measures, future observations should ensure that VLBI observations are more sensitive than the accompanying low resolution observations. However, the small bandwidths of current VLBI arrays makes this currently very time-expensive. For example, to achieve the same sensitivity as the current VLA observations of GOODS-N ($1.8\microJybm$) would require around 15.5 days with the VLBA (assuming 2\,Gbps data-rates and 10 telescopes)\footnote{These values were calculated using the \href{http://www.evlbi.org/cgi-bin/EVNcalc}{EVN calculator}.}. This overhead can be reduced, but with a reduction in the FoV, with the use of larger telescopes. For example, the EVN (with 2\,Gbps data-rates, 10 telescopes including the 100\,m Effelsberg and 76\,m Lovell telescope) requires just 3.15 days, albeit the effective field of view is much smaller at $8\farcm5$ compared to the 27\arcmin~of the VLBA. Despite the large time investments required at the moment, it is worth stressing that the operational costs of a VLBI array are far less than the cost of an IR satellite needed to provide radio-excess measurements. In addition, the proposed expansion to 4\,Gbps operations, the expanding number of VLBI ready telescopes, for example the East-Asian VLBI Network \citep{An2018} and the African VLBI Network \citep{Gaylard2011}, and the inclusion of ultra-sensitive elements such as MeerKAT and the SKA, will drastically reduce the integration time needed. 

Finally, it is worth stating that VLBI selects a `clean' sample of radio-selected AGN, that is, it is not affected by SF-related radio emission to a far lesser effect. The selection of a radio-AGN sample through radio excess measurements is a trade-off between contamination from star-forming galaxies, or a significant bias towards those objects with large radio excesses. \citet{Delvechhio2018} estimate that their 1$\sigma$ cut-off used in their radio-excess measurement means that up to a third of their sample could be contaminated by star-forming galaxies. This indicates that the combination of low and high resolution radio data is integral in order to select a clean, unbiased sample of radio-AGN. This will permit those sources with both star-formation and AGN present to be detected. This will be explored in a future $e$-MERGE collaboration paper.

\section{Conclusions}\label{Sec:P2-conclusions}

In this paper, we continued our analysis of the VLBI-selected AGN population introduced in \citetalias{Radcliffe2018:p1}. Using deep, multi-wavelength data in the GOODS-N field we conducted a test of the various AGN classification techniques, from X-rays to radio, in order to determine their performance on a VLBI-selected sample of AGN. Our conclusions are as follows.

We investigated 14 different AGN classification techniques that included, a) searching for MIR power-law AGN using the \textit{Spitzer} and WISE telescopes, b) X-ray AGN using deep 2\,Ms X-ray observations provided by the \textit{Chandra} telescope and c) radio-excess AGN which uses the far-IR-radio correlation to search for radio emission above what would be expected from star-formation alone. We find that \textbf{no} one single AGN classification technique can reliably identify all VLBI-selected sources. 

Both IR colour-colour methods and X-ray surveys are notably incomplete. In concordance with other studies, IR colour-colour methods only detect the most luminous AGN while, the latter fails due to a combination of sensitivity and Compton-thick sources. We performed stacking on the X-ray non-detections finding a soft hardness ratio suggesting that the lack of X-ray sensitivity is the main reason for the majority of the non-detections.

We find that the radio-excess parameter is the most reliable metric in identifying the majority of the VLBI-selected sample and that a combination of radio-excess, X-ray and IR can identify all VLBI-selected sources as AGN. This is verified by similar approaches used in other deep fields. Analysis of the entire VLBI-selected population revealed that around 42
\% of sources exist in intermediate redshift dust-poor galaxies with the majority having radiatively inefficient accretion upon their singular AGN classification through radio excess (and thus radio jets).

When extending the AGN classification methods to the total underlying AGN population, we find that VLBI and radio excess measurements are intrinsically linked and the differences between the number of AGN is related to sensitivity. With deeper VLBI observations, we would expect to detect the majority of the radio-excess selected sample. To conclude, milliarcsecond resolution VLBI remains the best method of identifying radio-AGN when ancillary IR data is unavailable. 

\begin{acknowledgements}\label{Sec:Acknowledgements}
	
	The authors extend our gratitude to the referee whose insightful and helpful comments improved this manuscript. The authors gratefully acknowledge Len Cowie and Amy Barger for allowing us to use unpublished data and would like to thank M. Mendez, R. Windhorst, J. McKean, P. Padovani and J. Hodge for their help. This research made use of Astropy, a community-developed core Python package for Astronomy \citep{Astropy,Astropy2018}. \\
	The research leading to these results has received funding from the European Commission Seventh Framework Programme (FP/2007-2013) under grant agreement No 283393 (RadioNet3). J.F.R. acknowledges the Science and Technologies Facilities Council (STFC), the Ubbo Emmius scholarship from the University of Groningen, and the South African Radio Astronomy Observatory (SARAO) whose funding contributed to this research. \\
	The European VLBI Network is a joint facility of independent European, African, Asian, and North American radio astronomy institutes. Scientific results from data presented in this publication are derived from the following EVN project code(s): EG078. $e$-MERLIN is a National Facility operated by the University of Manchester at Jodrell Bank Observatory on behalf of STFC. The National Radio Astronomy Observatory is a facility of the National Science Foundation operated under cooperative agreement by Associated Universities, Inc. This work is based on observations taken by the 3D-HST Treasury Program (GO 12177 and 12328) as well as GO 11600 and GO 13420 with the NASA/ESA Hubble Space Telescope, which is operated by the Association of Universities for Research in Astronomy, Inc., under NASA contract NAS526555. This publication makes use of data products from the Wide-field Infrared Survey Explorer, which is a joint project of the University of California, Los Angeles, and the Jet Propulsion Laboratory/California Institute of Technology, funded by the National Aeronautics and Space Administration. 
\end{acknowledgements}
\bibliographystyle{aa}
\bibliography{Nowhere_to_HideII}

\begin{appendix}

\section{Detailed descriptions of VLBI detected objects}\label{Appendix}
%

\subsection{Previous VLBI detections}\label{Appendix:Previous_detections}
These objects were previously detected by the \citet{chi2013deep} global VLBI observations. We have updated their descriptions using new data available for the GOODS-N field. As expected, we detect the majority (11/12) of the \citet{chi2013deep} sources. The only source not detected, J123642+621545, is discussed in Section~\ref{SSec:Non_detection:J123642+621545}.

\subsubsection{J123642+621331}\label{SSSec:123642+621331}
J123642+621331 lies in the \textit{Hubble} Flanking Fields (HFF). HST ACS optical-NIR imaging \citep{2004ApJ...600L..93G} reveal a very faint, red, disturbed galaxy. There seems to be some confusion in the literature regarding the redshift of this source and therefore an incorrect redshift is often used in subsequent analyses. Using the Keck telescope, \citet{waddington1999nicmos} detected a strong emission line that was interpreted as a $\mathrm{Ly\alpha}$ emission line, corresponding to a redshift of $z=4.424$. However, the line emission originates 1\farcs5 north-west from the radio/optical position of the galaxy. This was inferred to be caused by an in-falling gas cloud that was re-radiating $\mathrm{Ly\alpha}$ emission from the AGN into our line of sight. The optical-NIR continuum emission from HST ACS imaging casts doubt on this interpretation. If the source was at $z=4.424$, emission at the longest HST wavelengths would be blocked by the $\mathrm{Ly\alpha}$ forest. While very red, it is clearly observed in the F606W band, and still visible in the F435W band. In addition, based upon IR and UV properties, the object would have an exceptional star formation rate of $\sim 6300~\mathrm{M_\odot yr^{-1}}$ if it is at $z=4.424$ \citep{whitaker2014sfr}. Indeed, photometric estimates suggest that its redshift is more likely to be $\sim 2$. \citet{hasinger2008absorption} derive a redshift of $\sim1.77$, if the original $\mathrm{Ly\alpha}$ line is interpreted as a [O\textsc{II}]3727 line and \citet{Berta2011_photz} assigned a photometric redshift of $2.44\substack{+0.21 \\ -0.45}$ to this object. The more recent 3D-HST survey, \citet{Momcheva:2016fr} used HST/WFC3 grism spectral measurements to derive a redshift of $z = 2.012 \pm 0.002$ and a recent study by \citet{Murphy:2017ja} used a MOSFIRE K-band spectrum to derive a definitive redshift of 2.018 for this object.

\textit{Spitzer} IRAC ($3.5$, $4.5$, $5.6$, $8.0\micron$) and MIPS ($24\micron$) instruments detect a faint IR source suggestive of some star formation activity. However, the spectral index across the IRAC bands is $-0.89$, indicative of a hot dust component from additional AGN activity \citep{alonso2006infrared}. This source is also detected in the ISO $15\micron$ \citep{aussel1999}, \textit{Herschel} $100\micron$ and $160\micron$ PACS  \citep{Elbaz2011} and \textit{Chandra} X-ray 2\,Ms observations \citep{xue_xray_2016}. \citet{Murphy:2017ja} fitted to the SED of this object (excluding the radio flux densities) and found the stellar mass to be $M_{\ast} = 2.4 \times 10^{10}\,\mathrm{M_{\odot}}$, stellar mass fraction of 0.4, an IR luminosity of $2.3\times 10^{12}\,\mathrm{L_{\odot}}$, and a dust temperature of $70\,\mathrm{K}$. The remarkably high dust temperature indicates that the bolometric luminosity is dominated by AGN activity. 

Deep, combined $e$-MERLIN and VLA imaging \citep{2020MNRAS.tmp.1396M} and previous MERLIN and VLA imaging \citep{muxlow2005high,richards2000vla} indeed reveal a compact core with two-sided jet emission, further confirming the presence of an AGN. The object also exhibits a modest radio excess ($q_{24}=-0.39\pm0.03$). This is emboldened by previous VLBI surveys. \citet{garrett2001agn} EVN observations detect a compact core and \citet{chi2013deep} global VLBI observations show a core ($\sim 227\microJy$) and one-sided jet emission ($\sim120\microJy$). The \citet{chi2013deep} VLBI observations derived a total integrated flux of $\sim$ $350\microJy$ for this object and our new EVN observations only detect an integrated flux density of $233\pm27.9\microJy$. We note that this object is near the phase centre of the \citet{chi2013deep} and also these new observations, hence the primary beam correction cannot be responsible for this variability. The EVN VLBI observations show a slight hint of the counter jet as seen in the $e$-MERLIN observations. This is possibly due to the increased proportion of short baselines in the EVN-only observations, compared to the global VLBI array used in \citet{chi2013deep}. All of the evidence provided confirms that this source is a dust enshrouded nuclear starburst with AGN activity present.

\subsubsection{J123644+621133}\label{SSSec:123644+621133}

J123644+621133 is associated with a red elliptical galaxy ($I=21.6$) with a spectroscopic redshift of $z=1.0128$ \citep{Barger_specz_2008}. It is detected in IR \textit{Spitzer} IRAC and MIPS observations, but is not detected in any \textit{Herschel} bands \citep{Elbaz2011}. Based upon IR and UV fluxes, \citet{whitaker2014sfr} suggests that the host galaxy has a SFR of $\sim 4.5\,\mathrm{M_\odot\,yr^{-1}}$. 

The VLA, JVLA, MERLIN and $e$-MERLIN observe large scale radio structures with jets extending to $\sim 162\,\mathrm{kpc}$ from the central AGN bearing the signs of a classical FR-I radio galaxy. As a result, there is a very large radio excess of $q_{24} < -1.89$. The MIPS $24\micron$ flux is only $2.5\sigma$ detection, hence, it is an upper limit. Previous EVN 1.6\,GHz observations by \citet{garrett2001agn} revealed a compact core component and a 5$\sigma$ unresolved radio source located $\sim60\,\mathrm{mas}$ south of the core that coincides with the axis of the jet. Global VLBI \citep{chi2013deep} and these new EVN observations only detect a barely resolved, $411\pm44.7 \microJy$, compact radio core. The AGN core is a weak detection in the 2\,Ms \textit{Chandra} X-ray observations with an absorption-corrected luminosity of $1.1\times 10^{42}\,\mathrm{erg\,s^{-1}}$ \citep{xue_xray_2016}.

\subsubsection{J123646+621405}\label{SSSec:123646+621405}

J123646+621405 is associated with a face-on elliptical galaxy at $z=0.9610$ \citep{Barger_specz_2008}. The source is detected in all \textit{Spitzer} IRAC bands and 24$\micron$ MIPS. \citet{whitaker2014sfr} derives a strong SFR of $\sim 37.2\,\mathrm{M_\odot\,yr^{-1}}$. It is very luminous in X-rays \citep[$1.3 \times 10^{44}\,\mathrm{erg\, s^{-1}}$;][]{xue_xray_2016}, and exhibits a mild radio excess ($q_{24}=-0.13\pm0.04$) indicating AGN activity. VLA and $e$-MERLIN observations detect a compact core with two-sided emission overlying the nucleus of the host galaxy. 1.6\,GHz EVN observations of \citet{garrett2001agn} detect a $4\sigma$ radio source at this position. Global VLBI \citep{chi2013deep} detects a compact AGN core with an extension to the north, which is in concordance with the VLA and $e$-MERLIN morphologies. Our new EVN observations confirm this, also detecting the slight extension to the north. 

\subsubsection{J123653+621444}\label{SSSec:123653+621444}

J123653+621444 overlays the nucleus of an elliptical galaxy at $z=0.3208$ \citep{Barger_specz_2008}. It is detected in the \textit{Spitzer} IRAC bands and MIPS 24$\micron$. Based upon IR and UV properties it has a SFR of $\sim 0.7\,\mathrm{M_\odot\,yr^{-1}}$ \citep{whitaker2014sfr}. JVLA, $e$-MERLIN, VLA and MERLIN observations reveal a compact radio source with a jet extension to the east \citep{muxlow2005high}. The source shows a large radio excess ($q_{24}=-0.98\pm0.13$) and is seen to vary over time  \citep{richards1998vla,richards2000vla,morrison2010very}. Global VLBI observations detect a $80\microJy$ AGN core \citep{chi2013deep} and our 1.6\,GHz EVN observations also detect an AGN core with an integrated flux density of $\sim 117\microJy$. This is suggestive of an AGN being present. 

\subsubsection{J123608+621036}\label{SSSec:123608+621036}

J123608+621036 is hosted by a disturbed galaxy with a spectroscopic redshift of $z=0.679$ \citep{Barger_specz_2008}. The object is bright in the IR and has no radio excess ($q_{24}=0.98\pm0.03$) suggesting that the system is undergoing some merger-induced star formation. There is also AGN activity present. The object is detected in X-rays \citep{xue_xray_2016}, and combined MERLIN and VLA observations observe a compact core and two sided extended emission extending across the nucleus of the galaxy \citep{muxlow2005high}. Global VLBI observations detect a slightly extended, $140\microJy$, core \citep{chi2013deep}. Our EVN observations also detect a slightly extended radio core with a integrated flux density of 140 $\mu$Jy. 

\subsubsection{J123624+621643}\label{SSSec:123624+621642}

J123624+621643 overlays a red elliptical galaxy ($b=26.52$) with a spectroscopic redshift of $z=1.918$ \citep{Smail2004:sz}. $e$-MERLIN imaging reveals a compact component with a weak extension to the south-west that is in agreement with previous MERLIN-VLA observations \citep[][Muxlow et al. in prep.]{muxlow2005high,richards2000vla}. The object shows a large radio excess ($q_{24}=-1.08\pm0.08$) and was detected in global VLBI observations by \citet{chi2013deep}. 1.6\,GHz EVN observations also detect the milliarcsecond radio core.  Deep 2\,Ms {\it Chandra} X-ray observations \citep{xue_xray_2016} do not detect this object, which suggests that this system could harbour an obscured AGN.

\subsubsection{J123700+620910 / GN16}\label{SSSec:123700+620909 / GN16}
J123700+620910 has no optical counterpart in HST ACS imaging, but a faint, irregular host galaxy is present in HST NIR ($m_{\rm F125W} = 23.56$). This source has similar properties to J123642+621331. Deep \textit{Spitzer} IRAC and MIPS, \textit{Herschel} PACS and SPIRE and SCUBA $850\micron$ imaging \citep{Pope2006} detect this source, which is also known as the sub-mm galaxy GN16. Based upon IR and sub-mm observations, this object was classified as a starburst galaxy with a star-formation rate of $\sim1000\,\mathrm{M_{\odot}\,yr^{-1}}$ with a photometric redshift of 1.68. The 3D-HST survey \citep{Skelton_HST3D_2014} suggests that the redshift of this object is actually at $2.58\substack{+0.07 \\ -0.06}$ and \citet{whitaker2014sfr} derives a much more modest SFR of $598\,\mathrm{M_{\odot}yr^{-1}}$. The object is not detected in X-rays \citep{xue_xray_2016}, but exhibits a small radio excess ($q_{24}= -0.1\pm0.04$) which suggests there is a dust-obscured AGN present.

Indeed, MERLIN-VLA observations detects a compact core with an extension to the north-east \citep{muxlow2005high}. Global VLBI observed a $\sim150\microJy$ compact radio core \citep{chi2013deep} and our 1.6\,GHz EVN VLBI observes a $\sim163\microJy$ radio core with an extension to the north, that in agreement with MERLIN-VLA observations. This proves, without doubt, the presence of an AGN in this system. 

\subsubsection{J123715+620823}\label{SSSec:123715+620823}

J123715+620823 is associated with a faint, extremely red object ($R=24.25$ mag) with a spectroscopic redshift of $z=0.9335$ (Cowie 2016 priv. comm.). The source is faint in the \textit{Spitzer} IRAC bands ($\sim 15 \microJy$) and the $24\micron$ MIPS instrument. The object is luminous and resolved ($\sim1.89\,\mathrm{mJy}$) in recent VLA observations \citep{morrison2010very}, and has a relatively large radio excess value ($q_{24} = -0.99\pm0.06$). In the MERLIN $0\farcs2$ resolution imaging, the compact radio core is revealed and is unresolved. However, global VLBI observations resolves the core that shows extended emission to the north-east \citep{chi2013deep}. This source has experienced a considerable change in flux density between the 2004 global VLBI observations ($\sim0.65\,\mathrm{mJy}$) and these 2014 EVN observations ($\sim 2.8\,\mathrm{mJy}$). However, this source is 6\farcm4 from the \citet{chi2013deep} pointing centre so primary beam attenuation could contribute to this variability. 

\subsubsection{J123717+621733}\label{SSSec:123716+621733}

J123717+621733 is associated with a $I=22.2$ optical counterpart with a spectroscopic redshift of $z=1.146$ \citep{Barger_specz_2008}. HST NIR imaging shows an edge on spiral galaxy with a bright nuclear region. The object exhibits a high SFR of $\sim 316\,\mathrm{M_{\odot}\,yr^{-1}}$ \citep{whitaker2014sfr}. It has no $24\micron$ radio excess ($q_{24}=0.52\pm0.03$) but has a slight radio excess in the $100\micron$ radio excess measurement ($q_{100} = 1.01\pm0.05$).   
MERLIN-VLA observations detect a compact radio component with a 0\farcs6 one-sided extension to the south-west \citep{muxlow2005high}. New $e$-MERLIN observations suggest that there is an extension to the north-east \citep{2020MNRAS.tmp.1396M} with the emission extended across the face of the host galaxy, thus confirming that the origin of the radio emission is partially due to star-formation. Global VLBI observations detects a $177\microJy$ compact radio component \citep{chi2013deep}, and our 1.6\,GHz EVN observations detect a compact radio source with integrated flux density of $269 \pm 32.7\,\mu$Jy. This confirms the presence of an AGN in this system. \textit{Chandra} X-ray imaging categorise this as an AGN with a absorption-corrected flux of $1.5\times 10^{42}\ergs$ \citep{xue_xray_2016}.

\subsubsection{J123716+621512}\label{SSSec:123716+621512}

J123716+621512 is associated with a $I = 19.8$ elliptical galaxy at $z=0.5605$ \citep{Barger_specz_2008}. HST NIR imaging reveals that this system is undergoing a merger with a companion galaxy with a spectroscopic redshift of $0.559$ \citep{Skelton_HST3D_2014}. This corresponds to a merger separation of $13.7\,\mathrm{kpc}$. It is detected in all \textit{Spitzer} IRAC bands and the 24\,$\micron$ MIPS band. \citet{whitaker2014sfr} derive a SFR of $\sim 3.6\,\mathrm{M_\odot\,yr^{-1}}$. The source is very faint in wide-band X-ray observations ($3.7\times 10^{41}\,{\rm erg\,s^{-1}}$) and is not classified as an AGN based upon X-ray characteristics. The object has a mild radio excess, $q_{24}=-0.02\pm0.05$. MERLIN-VLA imaging shows a one-sided emission that extends $0\farcs 4$ to the south-east \citep{muxlow2005high}. Global VLBI observations detect a $\sim 100\microJy$ compact radio core and our new 1.6\,GHz EVN observations detects a compact radio core with an integrated flux density of $\sim 125\microJy$. This core has a slight extension to the north-east. This confirms the presence of a merger-induced AGN within this system.

\subsubsection{J123721+621130}\label{SSSec:123721+621130}

J123721+621130 is hosted by an extremely red ($K=21.06$ mag) galaxy with a photometric redshift of 2.02 \citep{Skelton_HST3D_2014}. \textit{Spitzer} IRAC and MIPS $24\micron$ along with \textit{Herschel} PACS and SPIRE bands detect a IR bright foreground galaxy approximately $3\farcs5$ to the NE. Due to the large IR resolutions, this makes any accurate IR measurement difficult for this source.  The source is detected in X-rays by \textit{Chandra} \citep{xue_xray_2016}.

MERLIN-VLA imaging reveals a compact core with one sided emission to the north \citep{muxlow2005high}. Global VLBI observations \citep{chi2013deep} detects a core and extension to the north that is in agreement with \citet{muxlow2005high}. Our EVN observations only detect the core but the integrated flux density ($364 \pm 41.6\microJy$) is significantly higher than the \citet{chi2013deep} observations who report a integrated flux density of $254 \pm 51\microJy$. However, this source is located $\sim5\arcmin$ from the pointing centre, therefore, the lack of primary beam correction in the \citet{chi2013deep} observations could cause this discrepancy.

\subsection{New VLBI detections}\label{Appendix:New_detections}
\subsubsection{J123555+620902}

J123555+620902 is associated with a red elliptical galaxy at a redshift of 1.8750 \citep{Barger_specz_2008} with a stellar mass of $10^{11.1}\,\mathrm{M_\odot}$ \citep{Hainline2011}. The source is a known sub-mm galaxy, and is detected in the SCUBA $850\micron$ survey, which estimates a dust temperature of $\sim47\,\mathrm{K}$ for this object \citep{Chapman2005:sm}. The source has a molecular gas mass of $10^{10.64\pm0.13}\,\mathrm{M_\odot}$ \citep{Bothwell2013}. The source shows signs of an AGN in the MIR (including WISE criteria), radio-excess, and is a bright X-ray AGN with an absorption corrected luminosity of $3.1\times10^{44}\,\mathrm{erg\,s^{-1}}$.  1.5\,GHz VLA observations show a barely resolved source with an integrated flux density of $192\microJy$ overlaying the optical maximum. These new VLBI observations confirm the AGN nature of this host with a primary beam corrected flux density of $100\microJy$. Due to the AGN signatures in all bands, this source is most likely a jetted RE-AGN whose AGN output dominates the bolometric luminosity of the object.

\subsubsection{J123607+620951}

J123607+620951 is hosted by a large edge-on spiral galaxy with a spectroscopic redshift of 0.6380 \citep{Barger_specz_2008} and a stellar mass of $10^{11.06}\,\mathrm{M_\odot}$ \citep{Miller2011}. A possible companion galaxy around 3\farcs5 from the VLBI host could be the cause of the enhanced AGN activity. However, redshift estimates for this galaxy, which are photometric only, are very uncertain. \citet{Skelton_HST3D_2014} and \citet{Yang_photz_2014} estimate a redshift of $0.6065$ and $0.565\substack{+0.052\\-0.067}$, respectively. 

This source shows evidence of AGN activity only in X-rays, and only has a very mild radio excess ($q_\mathrm{TIR} = 1.86\pm0.09$; see Figure~\ref{Fig:Radio excess q_TIR}). \citet{Simmons11} estimates that the central black hole has a mass of $10^{8.71\pm0.5} \,\mathrm{M_\odot}$ and is accreting at $\sim1.3\%$ of Eddington luminosity. The source is only slightly resolved in the 1.5\,GHz VLA observations with an integrated flux density of $205\microJy$. $e$-MERLIN observations show that the source is compact and point-like. These VLBI observations reveal an unresolved radio core with an integrated flux density of $118\pm21.2\microJy$. This detection confirms the underlying AGN nature of this source. 

\subsubsection{J123618+621541 / HDF130}\label{SSSec:123618+621541}
J123618+621541 is optically associated with a faint ($I=24.27$) elliptical galaxy at $z=1.993$ \citep{Smail2004:sz}. It is detected in \textit{Spitzer} IRAC and MIPS data, but is not detected in X-ray observations. It has a large radio excess value indicative of AGN activity ($q_{24}=-0.96\pm0.09$). This object is not detected in X-rays \citep{xue_xray_2016}. This object was originally categorised as a sub-mm faint star-forming radio galaxy because the UV spectra from the Keck/Low Resolution Imaging Spectrometer (LRIS) resembles that of a starburst galaxy \citep{Chapman2004:sm}. However, using high resolution MERLIN observations, \citet{Casey:2009gd} revealed that this galaxy is most probably a highly evolved giant elliptical galaxy, with a beamed low-luminosity AGN. The stellar mass and black hole mass is estimated to be $\sim 2.5\times10^{11} \,{\rm M_\odot}$ and $\sim 3.2\times10^{8}\,{\rm M_\odot}$, respectively. These evolved objects are rare at $z\sim2$, with the physical size of the host galaxy approximately 2\mbox{-}5 times larger than typical galaxies at the same redshift \citep{Casey:2009gd}. Our 1.6\,GHz EVN observations confirm the existence of an AGN with a detection of an AGN core with an integrated flux density of $192\microJy$. This was not detected by \citet{chi2013deep}, although being theoretically detectable (only 2\arcmin~from \citet{chi2013deep} pointing centre), thus suggesting that this source could exhibit some radio variability on milliarcsecond scales. 

\subsubsection{J123620+620844}\label{SSSec:123620+620844}

J123620+620844 is hosted by an elliptical galaxy ($I=21.05$) with a spectroscopic redshift of 1.0164 \citep{Barger_specz_2008}. It is detected in the \textit{Spitzer} IRAC and MIPS bands. \citet{whitaker2014sfr} derive a SFR of $\sim 3.5~\mathrm{M_\odot~yr^{-1}}$. The object exhibits a large radio excess indicating there is an active AGN ($q_{24} = -0.88\pm0.20$). High resolution, MERLIN-VLA imaging reveals a $115\microJy$ compact radio component with a small $\sim0\farcs2$ extension to the west. Our new EVN observations detect a $\sim 185\microJy$ radio core with a slight extension to the west, in agreement with the VLA-MERLIN observations. This source is identified as variable in \citet{Radcliffe2018:var}, further confirming the AGN nature, and the flux density differences between the EVN and VLA-MERLIN measurements. 

\subsubsection{J123621+621708}\label{SSSec:123621+621708}

Inspection of the HST F125W NIR images reveals two possible interacting galaxies that are 2\farcs73 apart. The VLBI source has a spectroscopic redshift of 1.988 \citep{Chapman2005:sm} whilst the possible companion has been prescribed the same redshift by \citet{Barger_specz_2008}. It is possible that these spectroscopic redshifts could be tied to the wrong galaxy because the \citet{Chapman2005:sm} survey uses the VLA radio positions as priors, where the radio emission is combined into one source. Photometric estimates put the VLBI host at $z\sim2.3$ and companion at a different redshift of $z\sim2.13$ \citep{Skelton_HST3D_2014}. However, the disturbed morphology does lend itself to an ongoing merger scenario. 

The VLBI-detection is weak ($\sim 135\microJy$) and both galaxies are detected in the 1.5\,GHz VLA observations. The VLBI-host and interacting galaxy have integrated flux densities of $143.4\pm14.7\microJy$ and $52.0\pm7.5\microJy$, respectively. The VLBI host exhibits a large radio excess ($q_{24} = -0.86\pm0.15$), while the companion exhibits no radio excess ($q_{24} = 0.83\pm0.21$). This indicates that the radio emission in the companion may be caused by merger-induced star-formation. Using the relation from \citet{Novak2017:co}, and assuming a spectral index of $-0.7$, we derive a SFR for the companion galaxy of $\sim 270\,\mathrm{M_\odot\,yr^{-1}}$.

\subsubsection{J123622+620654}\label{SSSec:123622+620654}

J123622+620654 is associated with a faint red host galaxy ($K = 22.59\,\mathrm{mag}$) with a photometric redshift of $1.937\substack{+0.120 \\ -0.117}$ \citep{Skelton_HST3D_2014}. It is detected in \textit{Spitzer} IRAC and MIPS instruments and the $160\micron$ Herschel. \citet{whitaker2014sfr} derive a SFR of $\sim 55~\mathrm{M_\odot yr^{-1}}$, indicating significant star-formation activity. The object exhibits a small radio excess ($q_{24} = -0.44 \pm 0.05$) and is not detected in X-rays \citep{xue_xray_2016}.

The object is present in re-processed MERLIN-VLA data, which reveals a compact radio core and a small two-sided extension in an east-west direction. Note that this was not reported in \citet{muxlow2005high} because the object is outside the survey range. Our EVN observations confirm the presence of an AGN, revealing a faint, compact radio core. 

\subsubsection{J123641+621833}\label{SSSec:123641+621833}

J123641+621833 is hosted by a spiral galaxy with a spectroscopic redshift of 1.1456 \citep{Barger_specz_2008}. It is detected in the \textit{Spitzer} IRAC and MIPS instruments. It is also detected in the \textit{Herschel} $100\micron$ and $160\micron$ bands. \citet{whitaker2014sfr} derives a SFR of $\sim 24.5~\mathrm{M_\odot\,yr^{-1}}$. The object has a large radio excess value ($q_{24} = -0.98\pm0.09$) and a \textit{Chandra} X-ray detection \citep{xue_xray_2016}, which suggests significant AGN activity. 

This object was out of the \citet{muxlow2005high} survey scope, but re-processed MERLIN-VLA data reveals an unresolved radio core. Our new EVN observations just resolve the object into a radio core and a small ($\sim$7 mas) extension to the west.

\subsubsection{J123650+620738}\label{SSSec:123650+620738}

J123650+620738 is associated with a disturbed irregular galaxy with a spectroscopic redshift of 1.6095. The object has a double nucleus that could be due to a dust band or an ongoing merger in the system. The VLBI detection is located in the south nucleus with a primary beam corrected flux density of  $98.7\pm19.9\microJy$. High resolution $e$-MERLIN observations (Muxlow et al in prep.) shows the radio emission is extended across both optical nuclei. This could be either due to jets originating from the AGN, starburst activity or a second AGN system. Closer inspection of the VLBI images reveal a possible $5.5\sigma$ peak located in the optical maximum of the north nucleus indicating that it could be another AGN. However, combined EVN $e$-MERLIN observations presented in future work should confirm this hypothesis. 

This object shows signs of AGN activity in all wavebands. The source is classed as a high-excitation emission line galaxy \citep{Trouille2008}, and shows evidence of power-law MIR AGN activity \citep{Donley2007:IR}. All other IR classification methods deem this an AGN and the object is extremely bright in the X-rays ($L_{0.5\mbox{-}7\,\mathrm{keV}} = 5\times10^{44}\,\mathrm{erg\,s^{-1}}$). Detailed SED fitting by \citet{DelMoro2016} show that 94\% of the rest-frame $6\micron$ emission is due to the AGN. Hence, the observed $24\micron$, corresponding to rest-frame $9.2\micron$ emission, is highly contaminated by AGN contributions. This is reflected in the $q_{24}$ parameter which shows no radio excess, but the longer wavelength estimates ($q_{100}$ and $q_\mathrm{TIR}$) indicating the presence of an AGN. 

\subsubsection{J123659+621833}\label{SSSec:123659+621833}

This object is associated with a faint, red host galaxy ($K=22.19$), with a photometric redshift of $2.167 \substack{+0.076 \\ -0.074}$ \citep{Skelton_HST3D_2014}. It is a faint detection in \textit{Spitzer} IRAC and MIPS and has a strong negative spectral index across the IRAC bands ($\alpha_{\mathrm{IR}} \sim -1.1$). This suggests that the MIR is dominated by AGN activity. The host galaxy is not detected in HST ACS observations \citep{2004ApJ...600L..93G}. \citet{whitaker2014sfr} derives a large SFR of $\sim 294~\mathrm{M_\odot yr^{-1}}$ for this object. It is also detected in \textit{Chandra} X-rays \citep{xue_xray_2016} and has a large radio excess ($q_{24} = -1.18 \pm 0.04$).

Re-processed MERLIN-VLA observations detects a bright compact object with a 1\arcsec~extension to the south. More recent VLA observations detects a 4\,mJy compact object and there is indications that the object could be variable. Our EVN VLBI observations prove unequivocally that an AGN is present. We reveal a 2.6\,mJy radio core which is extended in a north-west and south-east direction.  

\subsubsection{J123709+620838}\label{SSSec:123709+620838}

J123709+620838 is hosted by a $I=21.06$ galaxy with a spectroscopic redshift of $0.907$ \citep{Cowie2004z}. The object is detected in the \textit{Spitzer} IRAC and MIPS instruments with a derived SFR of $\sim 11.4~\mathrm{M_{\odot} yr^{-1}}$. The object has a mild radio excess ($q_{24}=-0.63\pm0.14$) and is detected in X-rays \citep{xue_xray_2016}.  The \textit{Herschel} SPIRE and PACS instruments also detect a source at this position, but it is uncertain whether the flux originates from the VLBI host galaxy. This is because the angular resolution is comparable to the separation of this source and a separate \textit{Spitzer}, radio and optical detection $\sim 4\farcs5$ away (J123709+620841). 

VLA observations suggests that this object is variable \citep{richards1998vla,morrison2010very}. Combined MERLIN-VLA data detect a compact radio core and faint extension to the west, indicative of AGN activity \citep{muxlow2005high}. Our 1.6\,GHz EVN observations reveal a compact object that has an extension to the west that is in agreement with MERLIN-VLA observations. 

\subsubsection{J123714+621826}\label{SSSec:123714+621826}

J123714+621826 is associated with a very faint, red galaxy ($K=24.41$ mag) with a tentative photometric redshift of $3.44$ \citep{Cowie2017:sc}. The object is faint  in \textit{Spitzer} IRAC and MIPS and is also detected by \textit{Herschel} PACS and SPIRE instruments. It has a large radio excess ($q_{24} = -1.04 \pm 0.06$) and is detected by the \textit{Chandra} X-ray telescope \citep{xue_xray_2016} indicative of an AGN. 

The object was not recorded in \citet{muxlow2005high}, but re-processed MERLIN-VLA data reveal a compact, 0.5 mJy radio core (not primary beam corrected). \citet{morrison2010very} VLA  and JVLA observations detect radio emission at a similar flux density ($\sim 0.6\,\mathrm{mJy}$) indicating that the majority of this radio emission occurs on sub-arcsecond scales. Our VLBI observations detect a $\sim 629\,\mathrm{\mu Jy}$ AGN core indicating that the radio emission of this object is AGN dominated.

\subsubsection{J123720+620741}\label{SSSec:J123720+620741}

J123720+620741 is associated with a $I = 20.4$ galaxy with a photometric redshift of $0.91 \substack{+0.05 \\ -0.03}$ \citep{Yang_photz_2014}. Radio observations indicate that this object is variable, confirming the presence of an AGN. \citet{richards1998vla} VLA observations in 1994 detected a $217\microJy$ radio source, while the 2006 VLA observation by \citet{morrison2010very} observed a 50\% decrease in flux density ($117\mathrm{\mu Jy}$). JVLA observations in 2011 saw only a small difference ($127.4\microJy$). Our new EVN observations detect a $\sim 112\microJy$ radio core and extension to the south, which suggests that the source may have increased in flux since 2011. 

\subsubsection{J123726+621129}\label{SSSec:123726+621129}

J123726+621129 is optically associated with a faint ($I=23.57\,\mathrm{mag}$) galaxy with a spectroscopic redshift of $1.2653$ \citep{Barger_specz_2008}, however Cowie priv. comm. suggests a spectroscopic redshift of 0.943 for this object. HST imaging reveals that the host galaxy morphology looks disturbed and may have undergone a recent merger \citep{2004ApJ...600L..93G}. The object is detected by all \textit{Spitzer} bands \citep{dickinson2003spitzer,ashby2013IRAC}, but is not detected by the \textit{Chandra} X-ray observatory \citep{xue_xray_2016}. \citet{whitaker2014sfr} derive a SFR of $\sim 17 ~\mathrm{M_\odot\,yr^{-1}}$, but this was assuming the photometric redshift of 1.67 from the 3D-HST survey \citep{Skelton_HST3D_2014}.

The VLA, JVLA, MERLIN and $e$-MERLIN observes a faint radio core and two-sided radio emission with a typical wide-angled tail (WAT) morphology. The radio lobes extend by $\sim 3\farcs8$ from the core in an east-west orientation. As a result, the object has a very high radio excess value $q_{24}=-2.29\pm0.15$. Our 1.6\,GHz EVN observations resolve out the vast radio lobes and hotspots, isolating the faint radio core, which is unresolved at a resolution of 16\,mas.

\subsubsection{J123649+620439}

J123649+620439 overlays the core of a bright elliptical galaxy ($I=17.1$) with a spectroscopic redshift of 0.1130 \citep{SDSS7:2009sd,Barger_specz_2008}. \textit{\textit{Chandra}} observations detect a faint X-ray source with an absorption corrected luminosity of $6.1\times 10^{40}\ergs$ \citep{xue_xray_2016}. VLA observations detect a $686\pm69\microJy$ object that is resolved, but has no morphological indication of AGN jet activity. The object shows no radio excess emission ($q_{24} = 0.35\pm0.06$), suggesting that some radio emission originates from star-formation related processes. Our EVN observations detect an AGN core, which is slightly resolved to the south-west. The AGN core has an integrated flux density of $102\microJy$. However, this source has not been primary beam corrected, because it lies outside the FWHM of the 100\,m Effelsberg telescope. 

\subsubsection{J123701+622109}\label{SSSec:123701+622109}

J123701+622109 overlies the nucleus of a $I=20.92\,\mathrm{mag}$ galaxy with a spectroscopic redshift of 0.8001 \citep{Barger_specz_2008}. The source is detected by \textit{Spitzer} IRAC and MIPS instruments and \citet{whitaker2014sfr} derives a SFR of $\sim 6.1\,\mathrm{M_\odot yr^{-1}}$. 
\citet{morrison2010very} VLA observations detect a 0.33\,mJy source at this position and re-processed VLA-MERLIN observations reveal an unresolved radio core, centred on the larger galaxy. The existence of an AGN is confirmed by our EVN observations that reveal a faint radio core. The object has a large radio excess value of $q_{24}=-1.31\pm0.19$. The AGN activity in this galaxy could have been induced by the interaction with the candidate companion galaxy. 

\subsubsection{J123739+620505}

J123739+620505 coincides with a very weak optical detection ($I=27.45$) and is not detected in \textit{\textit{Spitzer}} IRAC instruments \citep{Yang_photz_2014}. The object was assigned a very tentative redshift of $2.99\substack{+0.81 \\ -1.51}$. This object is out of the sky coverage of the majority of the GOODS-N multi-wavelength surveys, hence there are few multi-wavelength counterparts. VLA observations detect a barely resolved, $235\microJy$ radio detection. These VLBI observations detects a $194\microJy$ radio core, thus confirming the presence of an AGN in this system. We note that this source is not primary beam corrected. 

\subsubsection{J123751+621919}

J123751+621919 is hosted by an elliptical galaxy ($K_s = 20.294$) with a secure photometric redshift of $1.20\substack{+0.11\\-0.05}$ \citep{Yang_photz_2014}. This photometric redshift is in agreement with the ALAHAMBRA survey estimate of $1.239\substack{+0.039\\-0.049}$ \citep[][]{Molino2014}\footnote{Note these are 95\% confidence intervals}. The source is at the edge of the HST NIR (F125W and F160W) coverage, but is outside the field of view of the optical HST bands. The source has \textit{Herschel} counterparts, and the total radio excess measurement indicate the presence of an AGN ($q_\mathrm{TIR}=0.93\pm0.17$). X-ray and IR AGN measures show no sign of an AGN in this object. VLA radio observations reveal a partially resolved source with an integrated flux density of $155\microJy$. \citet{Radcliffe2018:var} reveals that the radio emission shows signs of variability, further strengthening the AGN nature of this object. VLBI observations show a resolved 181$\microJy$ core which has not been primary beam corrected. This indicates that the VLBI core flux density may be in excess of the VLA data presented in \citetalias{Radcliffe2018:p1}. This further reinforces the variable nature of this object, and the presence of an AGN in this system.

\subsubsection{J123523+622248}

J123523+622248 is hosted by a faint elliptical galaxy ($R=24.0$) with a tentative photometric redshift of $1.42\substack{+0.10\\-0.11}$ \citep{Wang2010,Yang_photz_2014}. The source is located outside the deep multi-wavelength coverage in GOODS-N, hence it has no HST or X-ray counterparts. The host has faint WISE counterparts in bands W1 and W2  \citep[$3.4\micron$ and $4.5\micron$ respectively;][]{Cutri2012:WISE}. However, the \citet{Stern2012:ir} AGN classification criterion does not classify this source as an AGN. These VLBI observations reveal the existence of an AGN, which is partially resolved, and exhibits a sub-jet to the south-west. Note that the source has no primary beam correction applied so no accurate flux density for this object can be obtained. VLA observations show a $1.7\,\mathrm{mJy}$ slightly resolved source which is extended in the same direction as the VLBI observations.

\subsubsection{J123510+622202}

J123510+622202 coincides with a faint, red galaxy ($V\approx26$) with an unknown morphological type \citep{capak2004HDFN,Wang2010}. Again, the source is located outside the deep multi-wavelength coverage therefore has no HST or X-ray coverage. As a result, the photometric redshift is fairly uncertain. We have adopted the \citet{Yang_photz_2014} value of $2.33\substack{+0.52\\-0.24}$ which is in agreement with the \citet{Rafferty2011} photometric redshift of $2.47\substack{+0.77\\-0.83}$. VLA 1.5\,GHz observations reveal an almost unresolved $1.2\,\mathrm{mJy}$ source. These VLBI observations reveal a compact unresolved core indicative of AGN activity. Again, the source has no primary beam correction applied so an accurate flux density cannot be obtained. 

\subsubsection{J123656+615659 - SDSS J123655.80+615659.3}

J123656+615659 is hosted by a compact elliptical galaxy ($g=21.7\,\mathrm{mag}$) with a spectroscopic redshift of 0.41938 \citep{Alam2015}. Unfortunately, this was overlooked at the time of publication of \citetalias{Radcliffe2018:p1} who quoted the photometric redshift of $0.39\substack{+0.05\\-0.04}$ from \citet{Yang_photz_2014}. This discrepancy makes little difference in the analysis presented here. The radio morphology of this object is a classic FR-I system ($P_\mathrm{1.4\,GHz} \approx 10^{25.1}$) with radio lobes extending to a projected linear size of approximately 100\,kpc. The total flux density is approximately 24\,mJy of which 7.2\% is polarised \citep[mainly the core and northern jet;][]{RudnickOwen2014}. The VLBI source is located well beyond the $25\,\mathrm{m}$ parabolic antenna FWHM, hence no primary beam correction has been attempted, but we note that the VLBI core flux density is in excess of $524\microJy$.

\begin{figure}
    \centering
    \includegraphics[width=\linewidth]{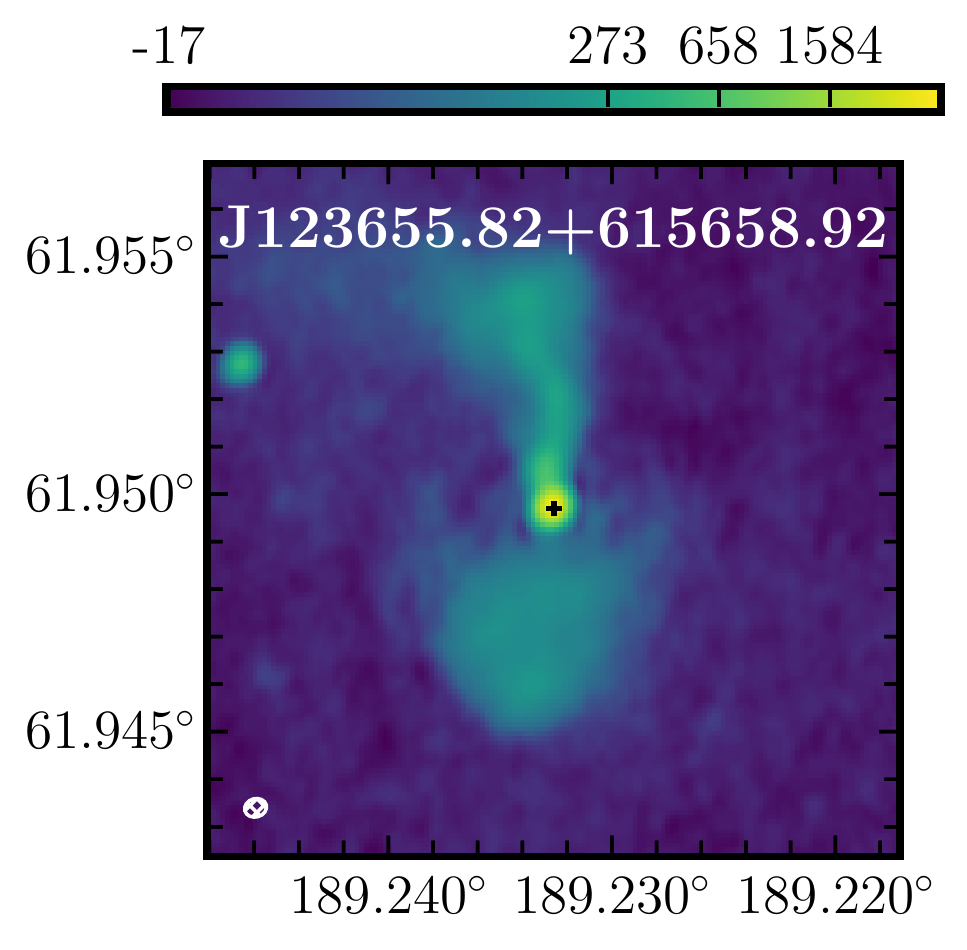}
    \caption[1.5\,GHz observations of FR-I source J123656+615659]{$1.5\,\mathrm{GHz}$ VLA observations of J123656+615659 showing a classical FR-I morphology. The units are in$\microJybm$}
    \label{fig:J123656_615659_radio}
\end{figure}

\subsection{Non-detection: J123642+621545}\label{SSec:Non_detection:J123642+621545}
\begin{figure}
    \centering
    \includegraphics[width=\linewidth]{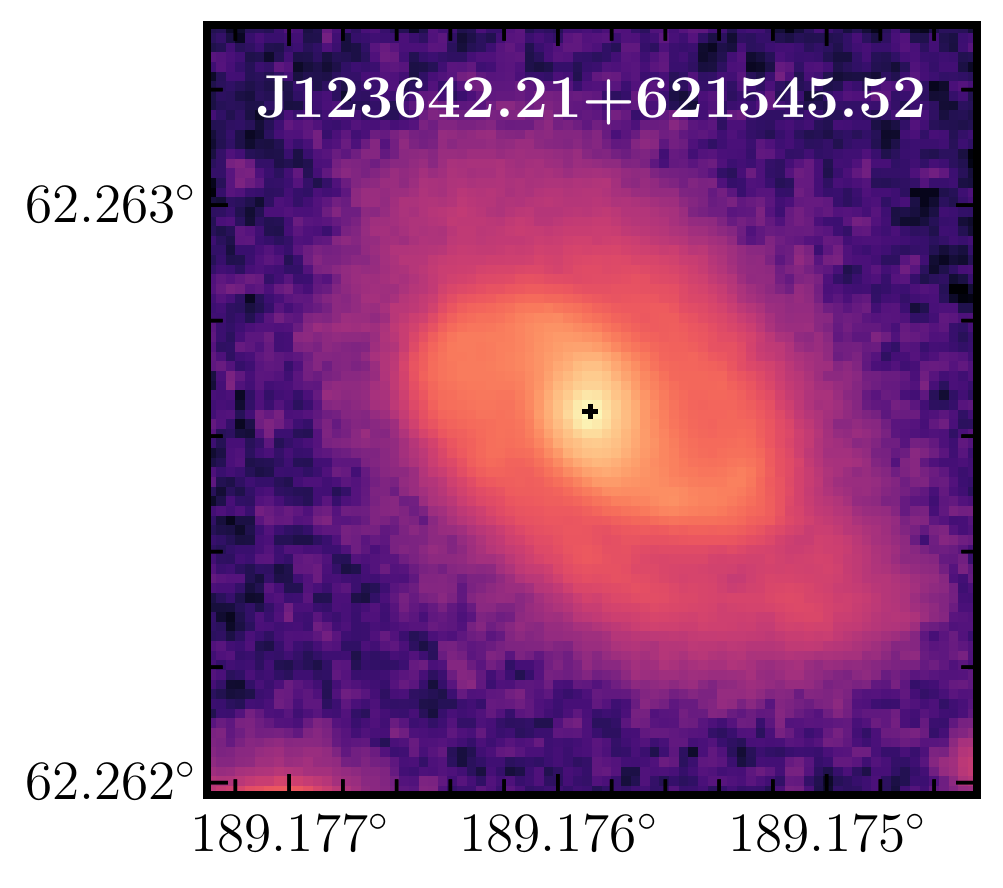}
    \caption[The host galaxy of VLBI variable source J123642+621545]{HST F160W imaging of the host galaxy of J123642+621545, a variable VLBI-selected source. This was detected in the 2004 Global VLBI observations \citep{chi2013deep}, but not in the 2014 observation presented in this paper. The quiescent radio emission is consistent with star-formation, that could indicate a nuclear starburst being present or low-luminosity AGN jet emission.}
    \label{fig:J123642+621545}
\end{figure}

Our observations did not detect this object, but \citet{chi2013deep} detected a $343\pm101\microJy$ radio core with a north-south extension. This indicates that there is a highly variable AGN present and is confirmed by the recent variability study presented in \citet{Radcliffe2018:var}. The \citet{chi2013deep} VLBI detection overlies the nucleus of a large spiral galaxy (see Figure\,\ref{fig:J123642+621545}) at $z=0.857$ \citep{Cowie2004z}. X-ray observations classifies this object as an AGN, with an absorption corrected $0.5\mbox{-}7\,\mathrm{keV}$ luminosity of $9.9\times10^{42}\,\mathrm{erg\,s^{-1}}$ \citep{xue_xray_2016}. $\mathrm{1.4\,GHz}$ 1996 VLA observations detect a flux density of $131\microJy$ and 2011 JVLA observations, at $\mathrm{1.5\,GHz}$, detected a similar flux of $\sim 150\microJy$. This is approximately $200\microJy$ less than the milliarcsecond scale core detected with global VLBI. The stable quiescent component emission overlays the nuclear region of the galaxy and falls on the radio-IR correlation ($q_{24} \sim 0.75$ and $q_{100}\approx 1.9$). This may indicative of nuclear star-formation, possibly induced by the AGN activity or vice-versa. UV and IR data estimate a SFR of $\sim 85.7\,\mathrm{M_\odot yr^{-1}}$ \citep{whitaker2014sfr}. From the low flux density state VLA emission, the radio SFR of this object is calculated to be $\sim 170~\mathrm{M_\odot yr^{-1}}$ using Eqn. 13 from \citet{Novak2017:co} and assuming a Chabrier IMF \citep{ChabrierIMF}. 

\end{appendix}

\end{document}